\pdfoutput=1
  \documentclass[10pt, logo, twocolumn, copyright]{nvidiatechreport}

  \usepackage[numbers,sort&compress]{natbib}
  \usepackage{fontawesome5}
  \let\monthyeardate\relax
  \newdateformat{monthyeardate}{\THEYEAR-\twodigit{\THEMONTH}-\twodigit{\THEDAY}}

\usepackage{xurl}
\usepackage{array}
\usepackage{colortbl}
\usepackage{multirow}
\usepackage{tikz}
\usepackage{algorithm}
\usepackage{algorithmic}
\usepackage{placeins}
\usepackage{needspace}
\usepackage{fancyhdr}
\usepackage{listings}
\definecolor{codekeyword}{HTML}{2B6A2E}
\definecolor{codehighlight}{HTML}{875400}
\definecolor{codecomment}{HTML}{5C5C5C}
\definecolor{codestring}{HTML}{2B5A96}
\newcommand{\codefontsize}{\fontsize{8}{10.25}\selectfont}
\lstdefinestyle{dcrcode}{
  language          = Python,
  deletekeywords    = [2]{id,apply},
  basicstyle        = \ttfamily\codefontsize,
  keywordstyle      = \color{codekeyword},
  commentstyle      = \color{codecomment},
  stringstyle       = \color{codestring},
  emph              = {send,stop_and_wait,cleanup,dispatch_hook},
  emphstyle         = \color{codehighlight}\bfseries,
  showstringspaces  = false,
  basewidth         = 0.55em,
  aboveskip         = 0pt,
  belowskip         = 0pt
}
\newcommand{\nowidows}{\clubpenalty=10000 \widowpenalty=10000 \displaywidowpenalty=10000 }
\newcommand{\nowidowpar}{{\nowidows\par}}
\nowidows%
\newcommand{\cmark}{\tikz[baseline=-0.55ex]{\filldraw[line width=0.45pt] (0,0) circle (0.72ex);}}
\newcommand{\xmark}{\tikz[baseline=-0.55ex]{\draw[line width=0.45pt] (0,0) circle (0.72ex);}}
\newcommand{\pmark}{\tikz[baseline=-0.55ex]{\draw[line width=0.45pt] (0,0) circle (0.72ex); \fill (0,0.72ex) arc[start angle=90, end angle=270, radius=0.72ex] -- cycle;}}
\definecolor{badgefill}{HTML}{262626}
\newcommand{\badgecirc}[4][0.25em]{\tikz[baseline={([yshift=-#1]c.center)}]{\node[fill=badgefill,circle,inner sep=0pt,minimum size=#2,text=white,font=\fontsize{#3}{#3}\selectfont\sffamily\bfseries](c){#4};}}
\newcommand{\myc}[1]{\badgecirc[0.32em]{8pt}{6pt}{#1}}
\newcommand{\stepn}[1]{\badgecirc[0.32em]{8pt}{6pt}{#1}}
\newcommand{\sysname}{\mbox{NeMo-DCR}}
\newcommand{\parhead}[1]{\noindent\textbf{\mbox{#1}}\hspace{1em plus 0.3em minus 0.2em}}
\newcommand{\sysfull}{\emph{\sysname} (Delta-Compressed Refit)}

\newcommand{\abstracttext}{%
Agentic reinforcement learning (RL) disaggregates training from rollout into different clusters, so each policy update must reach rollout clusters before the next batch: a full 1T checkpoint takes 87.5~min between two AWS regions, yet only 1--2\% of BF16 elements change their stored values per step. Concurrent systems use this sparsity but reimplement placement rules, assemble full tensors, rebuild values arithmetically, or rely on a cross-cluster collective, and none fully recovers from \mbox{mid-refit failures}. We present \sysfull{}, which sends only changes yet gives receivers the bits of a dense refit. \sysname{} describes changes in the \mbox{Hugging Face} canonical coordinates that both sides already share and leaves their placement to the serving runtime's native loader. It projects over 96\% of MoE weight bytes directly and converts the rest, under 4\%. It encodes changes as compressible XOR masks for over 96\% of these bytes and as overwrites where XOR masks cannot apply, streams them over object storage or a relay tree without a cross-cluster collective, and updates receivers in place, recovering from failures with overwrite retries and a joint commit. With real updates, \sysname{} is 7$\times$ faster than AReaL-DTE\@. In 3\% and 5\% stress cases, 30B--1T refits are 12--40$\times$ faster than a transport-only full-checkpoint reference over the same object storage, and would make a synchronous Kimi~K2 RL step \mbox{1.4--1.5$\times$} faster at 1T.
}

  \title{\sysname{}: Bit-Exact Delta-Compressed Refit for Scalable Agentic RL at Trillion-Parameter Scale}
  \author[1,2*]{Songlin Jiang}
  \author[2]{Zhiyu Li}
  \author[2]{Terry Kong}
  \author[2]{Yu Yao}
  \author[2]{Youngeun Kwon}
  \author[2]{\\Bernard Nguyen}
  \author[2]{Ashwath Aithal}
  \author[1]{Mario Di Francesco}
  \affil[1]{Aalto University}
  \affil[2]{NVIDIA}
  \correspondingauthor{}
  \hypersetup{pdftitle={\sysname{}: Bit-Exact Delta-Compressed Refit for Scalable Agentic RL at Trillion-Parameter Scale},
    pdfauthor={Songlin Jiang, Zhiyu Li, Terry Kong, Yu Yao, Youngeun Kwon, Bernard Nguyen, Ashwath Aithal, Mario Di Francesco},
    pdfkeywords={agentic reinforcement learning, reinforcement learning systems, weight synchronization, delta updates, distributed inference}}
  \makeatletter\g@addto@macro\ps@firststyle{\fancyfoot[R]{\footerfont\thepage}}\makeatother
  \renewcommand{\abscontent}{\noindent\parbox{\linewidth}{\absfont\theabstract}}
  \begin{abstract}
  \vspace{-12pt}\textbf{Abstract.} \abstracttext\par\vspace{2pt}
  \faBookOpen~\textbf{Documentation}: \href{https://docs.nvidia.com/nemo/rl/nightly/guides/refit.html}{NeMo RL weight refit guide}\quad
  \faGithub~\textbf{Code}: \href{https://github.com/NVIDIA-NeMo/RL/pull/2444}{NeMo RL pull request \#2444}\par\vspace{-7pt}
  \end{abstract}

\usepackage[varqu,varl]{inconsolata}
\hypersetup{bookmarksnumbered=true}
\usepackage[algo2e,ruled,vlined,linesnumbered]{algorithm2e}
\SetAlCapNameFnt{\small}
\SetAlCapFnt{\small}
\SetArgSty{textnormal}
\SetKwComment{Comment}{$\triangleright$\,}{}
\definecolor{algcomment}{gray}{0.25}

\SetCommentSty{commentfont}
\SetKwInOut{Input}{Input}
\SetKwInOut{Output}{Output}
\SetKwFor{For}{for}{}{}
\SetKwFor{While}{while}{}{}
\SetKwFor{ForAll}{forall}{}{}
\SetAlFnt{\small}
\SetKwIF{If}{ElseIf}{Else}{if}{}{else if}{else}{endif}
\SetAlgoSkip{}
\SetKwFor{Spawn}{spawn}{}{}
\SetKwBlock{Loop}{loop}{}
\SetKwBlock{Try}{try}{}
\SetKwFor{Catch}{catch}{}{}
\SetKw{Raise}{raise}
\SetKw{Await}{await}
\SetKw{WaitUntil}{wait until}
\SetKw{Pass}{pass}
\IncMargin{1em}
\theoremstyle{definition}
\newtheorem{proposition}{Proposition}
\ifdefined\pdfsuppressptexinfo\pdfsuppressptexinfo=-1 \fi
\begin{document}
\setlength{\parskip}{0.5\baselineskip plus 1pt}

  \maketitle

\section{Introduction}
\label{sec:intro}

Agentic\renewcommand{\thefootnote}{*}\begin{NoHyper}\footnotetext{Work conducted during an internship at NVIDIA.}\end{NoHyper}\renewcommand{\thefootnote}{\arabic{footnote}}{} reinforcement learning (RL) trains a policy on multi-turn interactions with tools and environments, collected in a stage called \emph{rollout}~\cite{gao2026rollart}. Rollout takes over 70\% of the wall-clock time and suits different hardware and parallel layouts than training, so systems disaggregate it onto separate clusters, each with its own parallel layout of \emph{ranks}, one process per GPU~\cite{gao2026rose,gao2026rollart}. Training and rollout then hold separate copies of the policy, so before each rollout batch, a \emph{refit} must deliver the updated policy to the \emph{receivers}, the rollout ranks of a serving runtime such as vLLM~\cite{kwon2023vllm}. Disaggregation also lets rollout run on idle serving GPUs~\cite{gao2026rose} or in other datacenters~\cite{zhong2025streamrl,ye2026tensorhub}, where every refit must cross a wide-area network.

\begin{figure}[!t]%
\centering
\includegraphics[width=\columnwidth]{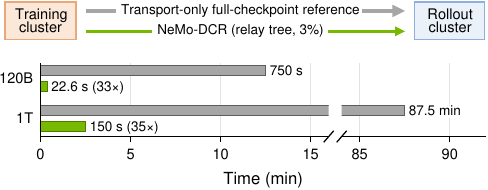}
\caption{\sysname{} relay-tree refits at a 3\% element change rate versus the transport-only full-checkpoint reference over object storage, with the speedup of each refit in parentheses (\S~\ref{sec:eval-scale}).}
\label{fig:densevsdelta}
\end{figure}

Nemotron-3 Super~\cite{nvidia2026nemotron3super} is an open 120B mixture-of-experts (MoE) model for agentic reasoning with a 247.2~GB checkpoint. Transferring it between two AWS regions through object storage takes 750~s, excluding save and load. The same transfer takes 87.5~min at 1T (Figure~\ref{fig:densevsdelta}), the scale of Kimi~K2~\cite{kimiteam2025kimik2}, an MoE model post-trained with agentic RL. Yet most of these bytes stay the same from one step to the next due to limited precision: RL fine-tuning widely adopts bfloat16 (BF16) for the weights that training updates and serving loads~\cite{qi2025fp16}. With only 8 significand bits, BF16 rounds most optimizer updates away, because a stored value changes only when an update crosses a rounding boundary. We quantified this in six widely used open-weight LLMs of different families and scales: per agentic RL step, only 0.6--1.2\% of their training-side source elements change value (Figure~\ref{fig:sparsity}). We call this fraction the \emph{element change rate}. A dense refit thus spends about 99\% of its \mbox{bytes on unchanged elements}.

\newcommand{\figsparsity}{%
\begin{figure}[!t]%
\centering
\includegraphics[width=\columnwidth]{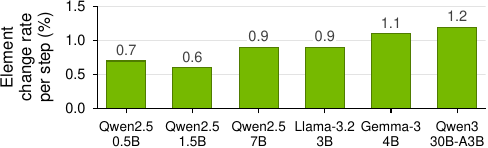}
\caption{Element change rate per step in six diverse open-weight LLMs, averaged over the first five steps: at most 1.2\%, so a dense refit spends about 99\% of its bytes on unchanged elements.}
\label{fig:sparsity}
\end{figure}}
\figsparsity%

\setlength{\parskip}{4pt plus 3pt minus 3pt}\setlength{\baselineskip}{12pt plus 0.3pt}One option to improve transfer efficiency is to send only changed values, namely a \emph{delta} between consecutive policy versions. This approach faces a challenge at every stage of a \mbox{refit}. The first is \emph{placement}: a receiver must know where and what to write, but sparse source changes do not map one to one onto its layout. Refitting may gather, split, or permute tensors between layouts~\cite{peng2026arealdte}, and a conversion can scale weights by a shared factor, so changing only that factor alters many converted values~\cite{nvidia2026megatronbridge}. Some systems reimplement the serving runtime's placement rules~\cite{gao2026rose,peng2026arealdte}, and others assemble and convert every changed tensor in full before extracting \mbox{a delta}~\cite{hu2026sparserlsync}.
Once placed, each update must be \emph{bit-exact}, because rollout--training mismatch destabilizes RL~\cite{qi2025fp16}. Arithmetic reconstruction can introduce rounding errors~\cite{peng2026arealdte}, while absolute overwrites avoid such errors~\cite{miahi2026pulse} but resend unchanged bits within changed values.
Exact payloads then need efficient \emph{delivery}: their volume must scale with the changed set, and delivery must run without a cross-cluster collective and overlap the other refit stages. A cross-cluster collective~\cite{nemorl2026reshard}, such as an NCCL broadcast, couples receiver membership to the training cluster and adds timeouts on failure. The next rollout waits for the whole update, so serial construction and application add to transfer time \mbox{and reduce the savings}.
Finally, applying updates in place avoids a receiver-side baseline copy but complicates \emph{recovery}: an interrupted refit can leave a mixture of old and updated values that matches neither policy version. Because a delta can be applied only to the version it was computed from~\cite{miahi2026pulse}, recovery must complete the refit before rollout resumes and keep receivers aligned with the source baseline \mbox{that the next refit compares against}.

The six requirements of \S~\ref{sec:goals} detail these challenges, and no state-of-the-art system meets all of them (Table~\ref{tab:goals}). We close this gap with \emph{bit-exact delta-compressed refit} (\sysname{}), which sends only changes yet gives every receiver the same parameter and buffer bits as a dense refit of all weights. The key idea is to describe changes in \emph{canonical coordinates}, the tensor names and indices of the Hugging Face checkpoint format that both training and serving already support. Because each side already maps its own layout to and from this format, no pair of training and serving layouts needs its own \mbox{delta mapping}. Fixed affine index mappings project most changes into these coordinates, and residual conversion handles the rest. The serving runtime's native loader maps these coordinates to receiver storage, as a page table maps virtual to physical addresses, so \sysname{} needs no model-specific placement logic of its own.
\begin{table*}[!t]
\centering
\footnotesize
  \caption[Refit approaches assessed against the six requirements]{Refit approaches assessed against the six requirements of \S~\ref{sec:goals}: \protect\cmark{} met, \protect\pmark{} partly met, and \protect\xmark{} not met or not addressed in the cited source. Appendix~\ref{app:comparison} details the assessment of each system and discusses additional related work.}
\label{tab:goals}
\renewcommand{\arraystretch}{0.934}
\setlength{\tabcolsep}{3pt}
\begin{tabular}{@{}>{\raggedright\arraybackslash}p{0.24\textwidth}>{\raggedright\arraybackslash}p{0.215\textwidth}cccccc@{}}
\toprule
 & & G1 & G2 & G3 & G4 & G5 & G6 \\
\hphantom{\myc{A}} \textbf{Approach} & \textbf{Systems} & \textbf{Native} & \textbf{Bit-exact} & \textbf{Delta} & \textbf{Decoupled} & \textbf{Pipelined} & \textbf{Recovery} \\
\midrule
\smash{\myc{A}} Dense collective refit & NCCL-Reshard~\cite{nemorl2026reshard} & \pmark & \cmark & \xmark & \xmark & \pmark & \pmark \\
 & checkpoint-engine~\cite{moonshot2026checkpointengine} & \cmark & \cmark & \xmark & \xmark & \pmark & \pmark \\
\arrayrulecolor{black!25}\midrule\arrayrulecolor{black}
\smash{\myc{B}} Full-checkpoint transfer & TensorHub~\cite{ye2026tensorhub} & \pmark & \cmark & \xmark & \cmark & \xmark & \pmark \\
 & ByteCheckpoint~\cite{wan2025bytecheckpoint} & \xmark & \cmark & \xmark & \cmark & \cmark & \pmark \\
\arrayrulecolor{black!25}\midrule\arrayrulecolor{black}
\smash{\myc{C}} Arithmetic reconstruction & ROSE~\cite{gao2026rose} & \xmark & \xmark & \cmark & \cmark & \cmark & \xmark \\
 & AReaL-DTE~\cite{peng2026arealdte} & \pmark & \xmark & \cmark & \cmark & \xmark & \pmark \\
 & AuroraRL~\cite{ruan2026aurorarl} & \xmark & \xmark & \cmark & \cmark & \cmark & \pmark \\
\arrayrulecolor{black!25}\midrule\arrayrulecolor{black}
\smash{\myc{D}} Absolute overwrites & PULSE~\cite{miahi2026pulse} & \xmark & \pmark & \cmark & \cmark & \pmark & \pmark \\
 & SparseRL-Sync~\cite{hu2026sparserlsync} & \cmark & \pmark & \cmark & \xmark & \xmark & \xmark \\
 & verl~\cite{verl2026deltasharded} & \cmark & \pmark & \cmark & \xmark & \cmark & \xmark \\
\arrayrulecolor{black!25}\midrule\arrayrulecolor{black}
\smash{\myc{E}} Checkpoint patching & slime~\cite{thudm2026slime} & \cmark & \cmark & \cmark & \cmark & \xmark & \pmark \\
\midrule
\hphantom{\myc{A}} Bit-exact delta refit & \textbf{\sysname{}} (this paper) & \cmark & \cmark & \cmark & \cmark & \cmark & \cmark \\
\bottomrule
\end{tabular}
\end{table*}
Its contributions form one design built on four principles, one per challenge.
\begin{enumerate}[label=\textbf{P\arabic*},leftmargin=*]\addtolength{\itemsep}{1.3pt}%
  \item \textbf{Placement through shared coordinates.} Each shard has one \emph{owner}, the training rank that detects changed elements and projects affine changes into canonical coordinates for over 96\% of the weight bytes in a Qwen3 and a Nemotron MoE checkpoint. Residual conversion compares converted tensors with a residual baseline for the rest and for changes of shared scale factors. Skipping full-tensor assembly and conversion speeds up Qwen3 delta construction by \mbox{1.08--1.16$\times$} \mbox{with the same overwrite encoding (\S~\ref{sec:bdt})}.
  \item \textbf{Exactness through bitwise \mbox{encoding}.} XOR masks encode affine changes whose projection and loader preserve stored bits and that no other write overlaps. They cover over 96\% of MoE weight bytes, and overwrites encode the rest. Both are exact, and because XOR masks compress better, mixed encoding cuts Qwen3 payload bytes by 38--40\% versus overwrites at 3\% and 5\% change rates \mbox{(\S~\ref{sec:codec})}.
  \item \textbf{Delivery by streaming compressed payloads.} Owners send payloads over object storage or a relay tree, without a cross-cluster collective, while constructing the rest of the delta. In synchronous RL, receivers apply payloads as later ones arrive. In asynchronous RL, rollout serves until the whole delta is staged and pauses only while receivers apply it and owners update the baseline. In Nsight Systems traces of 30B--550B relay-tree refits at 3\% and 5\%, delta construction overlaps transport, which takes at least 77\% of refit \mbox{latency (\S~\ref{sec:protocol})}.
  \item \textbf{Recovery through overwrite retries.} Receivers intercept the native loader's final GPU storage copies and apply updates in place, with no receiver-side baseline copy. Retries overwrite the complete changed set, repairing partial writes, and new receivers can join between attempts. After all current receivers apply the updates, a joint commit binds the policy to its source baseline, which owners then update in place. Proposition~\ref{prop:bdt} and its extension to retries guarantee bit-exactness. With receivers killed mid-refit in asynchronous Qwen3 runs, \sysname{} training aligns with the mean reward and KL trajectories of dense \mbox{NCCL refits} without kills \mbox{(\S~\ref{sec:apply})}.
\end{enumerate}

Together, the four principles make \sysname{} 7.0--7.3$\times$ faster than \mbox{AReaL-DTE}~\cite{peng2026arealdte} on two Qwen3 models with real updates that change about 2\% of elements per step. In 3\% and 5\% stress cases, which exceed every rate in Figure~\ref{fig:sparsity}, \mbox{30B--1T} refits are \mbox{12--40$\times$} faster than a transport-only full-checkpoint reference over the same object storage, and 1T refits would make a synchronous Kimi~K2 RL step \mbox{1.4--1.5$\times$} faster than with this reference (\S~\ref{sec:evaluation}).

\setlength{\parskip}{0.5\baselineskip plus 1pt}\newpage\normalsize%
\section{Motivation and Requirements}
\label{sec:background}

\subsection{Refit across layouts and versions}
\label{sec:boundary}

\begin{figure}[!htbp]%
\centering
\includegraphics[width=\columnwidth]{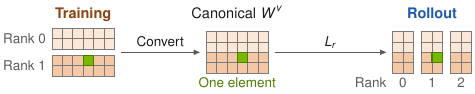}\\[5pt]
\includegraphics[width=\columnwidth]{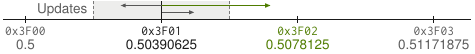}
\caption{One tensor in the training, canonical, and rollout layouts (top). Only updates that cross a BF16 rounding boundary change the stored bits (bottom).}
\label{fig:layouts}\label{fig:bf16round}
\end{figure}

Training systems and serving runtimes store the same policy in different layouts. Training systems shard parameters, gradients, and optimizer state to fit in GPU \mbox{memory} while maximizing update throughput~\cite{narayanan2021megatron,rajbhandari2020zero}. Serving runtimes organize weights for inference and maintain request state such as a paged key-value (KV) cache~\cite{kwon2023vllm}. RL systems that coordinate the training and rollout sides must bridge the two layouts~\cite{sheng2025hybridflow,hu2025openrlhf,jiang2026nereus}. The two sides choose their parallel degrees independently, so one training shard can overlap several rank-local slices on receivers, and one slice can combine parts of \mbox{several shards (Figure~\ref{fig:layouts}, top)}.

Both layouts map to canonical tensors: the model's tensors in the Hugging Face checkpoint's naming and layout, independent of either cluster's sharding. Let $W^v$ denote the canonical tensors at version $v$, which hold every parameter and buffer of the model. A receiver $r$ stores $W^v$ through its native loader $L_r$, the serving runtime's weight loader, which may split, fuse, pad, skip, transpose, or cast tensors before writing receiver storage $P_r^v=L_r(W^v)$. For~\mbox{example}, vLLM's loader fuses query, key, and value projections and splits routed experts across ranks. A \emph{loader path} is the sequence of operations that a loader applies to canonical tensors, such as these fusions and splits. We consider paths in which each stored element comes from at most one element of the canonical tensors, called its \emph{canonical element}.

A refit must bring each receiver to $P_r^{v+1}=L_r(W^{v+1})$, the result of a \emph{dense refit}. Approaches A and B of \mbox{Table~\ref{tab:goals}} send every weight to produce it. We call each refit to version~$v{+}1$ a \emph{transition}, which may take several attempts. The receivers that an attempt updates are its \emph{required receivers}.

\subsection{Few BF16 values change per step}
\label{sec:sparsity}

BF16 rounding absorbs most updates: in Figure~\ref{fig:bf16round}, bottom, all values in the shaded interval round to \texttt{0x3F01}. Figure~\ref{fig:sparsity} covers the first five steps of Group Relative Policy Optimization (GRPO)~\cite{shao2024deepseekmath} under an agentic software-engineering (SWE) RL recipe~\cite{nemorl2026sweqwen} with AdamW at a constant $10^{-6}$ learning rate and no weight decay. Other reports give change rates of 1.5--2.0\% in Qwen3 RL~\cite{peng2026arealdte} and average deltas of 1.98\% of a 1{,}024~GiB checkpoint~\cite{fireworks2026composer}.

\subsection{Requirements and related work}
\label{sec:goals}

To exploit this sparsity, we refine the four challenges of \S~\ref{sec:intro} into six requirements (Table~\ref{tab:goals}). Placement logic outside the serving runtime breaks with each model or runtime version, so updates belong where the native loader would place them, using only existing conversion rules (\textbf{\mbox{G1 Native}}). Each delta builds on the bits the last one left, so a rare error persists into every later version. Exactness therefore needs a proof that a delta refit from receiver storage $P_r^v$ yields the parameter and buffer bits of a dense refit (\textbf{\mbox{G2 Bit-exact}}). For delivery, savings come when transfer volume scales with the update: the changed values, their locations, and payload metadata (\textbf{\mbox{G3 Delta}}). A cross-cluster collective would let a failure or membership change on one side stall the other, so delivery avoids one (\textbf{\mbox{G4 Decoupled}}). Serial refit stages would offset the savings, so delivery overlaps them (\textbf{\mbox{G5 Pipelined}}). A delta fits only the version it was computed from, so recovery must ensure that a repeated or interrupted delivery leaves no trace (\textbf{\mbox{G6 Recovery}}). A duplicate or an unrepaired partial write would leave receivers between versions. If the policy advanced before every required receiver held $P_r^{v+1}$, or without the source baseline, the next delta would not match receiver storage.

Other refit systems send every weight~\cite{jaghouar2025intellect2,jiang2026weightbridge}, rebuild the full model from deltas~\cite{cursor2026composer,fireworks2026composer,huggingface2026trldelta}, or apply sparse overwrites in place without recovery~\cite{vllm2026sparse}. Beyond refits, related techniques include gradient compression~\cite{lin2018dgc,vogels2019powersgd}, low-rank adaptation~\cite{hu2022lora}, delta file transfer and lossless weight compression~\cite{tridgell1996rsync,hershcovitch2025zipnn}, and compression of fine-tuning deltas and checkpoints~\cite{liu2024bitdelta,li2024excp,yao2025deltazip,eisenman2022checknrun}.

\section{Design Overview}
\label{sec:design}
\label{sec:overview}
Training shards and receiver storage use different layouts, which depend on parallel degrees and on how each runtime fuses, pads, and slices tensors, so a delta needs a description that is independent of both layouts. Both sides already share one: the Hugging Face checkpoint that each imports and exports. A change described in these canonical coordinates means the same to every receiver, whatever its layout, so one payload serves all of them, and owners need no knowledge of receiver storage. The principles of \S~\ref{sec:intro} shape the refit of Figure~\ref{fig:architecture}, where a \emph{coordinator} drives each attempt.

\newcommand{\figarchitecture}{%
\begin{figure}[!t]%
\centering
\includegraphics[width=\columnwidth]{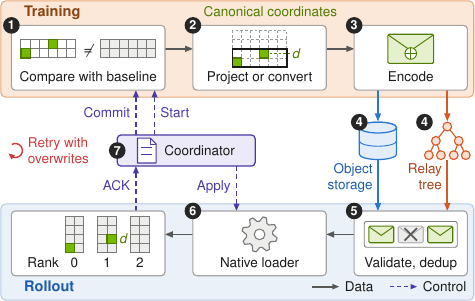}
\caption{\sysname{} architecture and the numbered steps of one refit, with changed elements in green.}
\label{fig:architecture}
\end{figure}}

Owners detect changes and project them into canonical coordinates (\stepn{1}--\stepn{2}, P1, \S~\ref{sec:bdt}), encode them as XOR masks or overwrites (\stepn{3}, P2, \S~\ref{sec:codec}), and stream compressed payloads to rollout nodes (\stepn{4}, P3, \S~\ref{sec:protocol}). Receivers validate and apply them in place through the native loader, and a joint commit advances policy and source baseline together, with overwrite retries after failures (\stepn{5}--\stepn{7}, P4, \S~\ref{sec:apply}).\nowidowpar

\parhead{Example.}Suppose the BF16 element of Figure~\ref{fig:layouts}, top, changes from \texttt{0x3F01} to \texttt{0x3F02} in a matrix sharded by rows in training and by columns in rollout. Its owner detects the change (\stepn{1}), projects it to canonical destination~$d$ (\stepn{2}), and emits the XOR mask \texttt{0x0003} (\stepn{3}) to every rollout node (\stepn{4}). Receivers validate the mask (\stepn{5}), and each one whose column slice holds~$d$ XORs it into \texttt{0x3F01} at the loader's final GPU storage copy (\stepn{6}) and acknowledges completion (ACK). The joint commit publishes the policy, and the owner's baseline becomes \texttt{0x3F02} (\stepn{7}). If the attempt fails midway, the retry overwrites~$d$ with \texttt{0x3F02} wherever it is stored, whatever \mbox{the earlier writes did}.

\figarchitecture%

\parhead{Implementation.}\sysname{} is about 7K lines of Python on top of Megatron Bridge conversion~\cite{nvidia2026megatronbridge} and the native loader of vLLM~0.17.0~\cite{kwon2023vllm}. It hooks into each at one point: the conversion's mapping classes, which decide which changes an owner can project directly, and the loader's final GPU storage copies, where \mbox{receivers} filter writes. Other stacks offer the same two hooks: \mbox{Megatron-style} splits~\cite{narayanan2021megatron} and ZeRO's flattened partitions~\cite{rajbhandari2020zero} map shards to canonical tensors by fixed \mbox{indices}, and SGLang~\cite{zheng2024sglang}, like vLLM, loads them through per-parameter \mbox{weight loaders~\cite{sglang2026loader}}.

\FloatBarrier
\section{Direct Projection and\texorpdfstring{\\*}{ }Residual~Conversion}
\label{sec:bdt}

Following P1, owners emit only changes (G3) and address each in canonical coordinates, so that the native loader can place it (G1). Two alternatives fall short: assembling and converting every changed tensor before extracting a delta~\cite{hu2026sparserlsync,thudm2026slime} gathers full tensors on every refit, and reimplementing placement for receivers~\cite{gao2026rose,peng2026arealdte} ties the refit system to each model and runtime version. \sysname{} projects most changes directly from shards and converts \mbox{the rest}.

The changes that an owner can directly project depend on the \emph{conversion tasks} through which the training framework, here Megatron Bridge, converts training shards to canonical tensors~\cite{nvidia2026megatronbridge}. \emph{Affine} tasks convert each source element independently and deterministically with the fixed affine index \mbox{mappings} of supported classes, such as those for replicated and row- or column-split shards. All other tasks, such as those for fused query/key/value (Q/K/V) source tensors, are \emph{residual} and require constructing canonical tensors. Affine mappings cover 96.3\% and 97.0\% of the weight bytes in the Qwen3-30B-A3B and \mbox{Nemotron-3-Ultra-550B-A55B} MoE checkpoints, respectively. The residual tensors are the attention Q/K/V tensors, the embedding and output head, and, in Nemotron, the input projections and convolutions \mbox{of its Mamba layers}.

\subsection{Ownership and change detection}
\label{sec:detect}

Data-parallel replicas stay bitwise identical, so one owner compares each unique source element with the source baseline. A stable hash of each shard's name picks its owner among the replicas, which balances ownership among them and keeps it fixed across attempts. Owners detect changes against the committed source baseline $B^v$, which holds a \emph{task tracker} with the \mbox{version-$v$} values of each task's training shard and a \emph{residual baseline} of canonical tensors for residual tasks. For task $T$, let $S_T^{v+1}$ be its \emph{candidate} shard of updated values, $A_T^v$ its tracker, and $J_T$ its owned indices. Let $I(X)_j$ be the stored bit pattern of $X$ at $j$. The changed indices $C_T$ are the $j\in J_T$ with \mbox{$I(S_T^{v+1})_j\ne I(A_T^v)_j$}. Comparing stored bit patterns catches every change, even when a zero changes sign.

\parhead{Owner memory.}Owners keep the source baseline in host memory, so it takes no GPU memory from training. Because each unique source element has one owner, the trackers together hold one copy of the source weights. For the \mbox{Qwen3-30B-A3B} and \mbox{Nemotron-3-Ultra-550B-A55B} checkpoints, the residual baseline adds 3.69\% and 3.01\%, so the source baseline totals about 1.04$\times$ and 1.03$\times$ their respective sizes. Balanced ownership spreads it across the training ranks: for the 550B checkpoint, each of the 32 ranks of \S~\ref{sec:eval-setup} holds \mbox{about 36~GB}.

\subsection{Direct projection}
\label{sec:projection}

Most conversions just move elements. For such an affine task, shard geometry fixes the canonical destination of each change, so an owner addresses its changes without assembling any tensor. For a changed index $j$ of affine task $T$, the index mapping $\pi_T$ identifies the canonical destination, and the source projection $Q_T$ \mbox{produces its updated value}:
\begin{equation}
 d = \pi_T(j), \qquad
 o_T^{v+1}(j) = I(Q_T(S_T^{v+1}))_d.
 \label{eq:affine}
\end{equation}
Here, $Q_T$ maps a shard into canonical coordinates, and $d=(n,k)$ is a tensor name and \mbox{flat index}.

The owner emits each change once for all receivers without knowing their storage offsets. In the \mbox{example} of \S~\ref{sec:design}, a changed local element $(i,y)$ of the row shard that begins at row $i_0$ maps to canonical destination $(n,k)$, where ${k=(i_0+i)w+y}$ for $w$ columns (Figure~\ref{fig:residual}, top). Each receiver's native loader then \mbox{selects its own column slice}.

\newcommand{\figresidual}{%
\begin{figure}[!t]%
\centering
\includegraphics[width=\columnwidth]{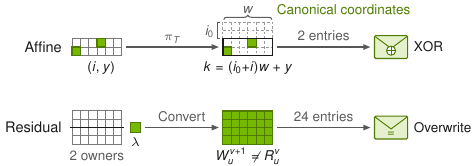}\\[17pt]
\includegraphics[width=\columnwidth]{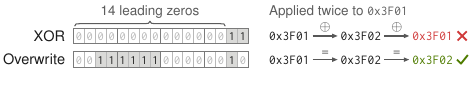}
\caption{Affine changes on representation-preserving paths yield XOR entries, and a changed scale factor~$\lambda$ yields an overwrite entry for every value it scales (top). Applied twice, an XOR entry restores the old bits, and an overwrite entry keeps the updated bits (bottom).}
\label{fig:residual}\label{fig:xorbits}
\end{figure}}
\figresidual%

\subsection{Residual conversion}
\label{sec:residual}

A residual task lacks a supported mapping, combines several owners' inputs into one value, or spreads a source change across several outputs, as in stacking, vocabulary padding, tied weights, adapters, postprocessing, and quantization scales. \sysname{} therefore constructs the canonical tensors that such tasks affect and compares their bits with the residual baseline (Figure~\ref{fig:residual}, top).

To find the affected tensors, change detection covers every input of a conversion. A fixed rule may multiply a shared scale factor~$\lambda$ into each source value before a cast, so a change in $\lambda$ alone alters many converted values. Each owner flags the residual tasks whose shards changed, and an all-reduce within the training cluster combines the flags by task. Through known conversion dependencies, the flagged tasks mark every canonical tensor they affect directly or indirectly, forming the \emph{required tensor set} $\mathcal{U}_{\mathrm{req}}$.

For each required tensor $u$, every owner converts its part. The owner picked by a hash of $u$'s name assembles the parts into~$W_u^{v+1}$ and collects the set $D_u$ of canonical destinations whose bits differ from the committed residual baseline~$R_u^v$. Residual tensors outside $\mathcal{U}_{\mathrm{req}}$ keep their committed values.

Direct projection and residual conversion feed one changed set: the emitted destinations combine direct projections $\pi_T(C_T)$ and residual changes $D_u$. Each emitted destination has a \emph{delta record} with its updated value, and these records form the complete changed set~$\mathcal{E}$ that \S~\ref{sec:codec} encodes.

\section{Mixed XOR/Overwrite Encoding with~Compression}
\label{sec:codec}
\label{sec:codec-repr}

Following P2, \sysname{} rejects arithmetic deltas: for an old value~$\alpha$ and an updated value~$\beta$, rounding can make ${\alpha+(\beta-\alpha)}$ differ from $\beta$. \sysname{} sends XOR masks of stored bits or absolute overwrites, which both reproduce the updated bits exactly (G2): a mask flips each bit that differs, and an overwrite replaces the stored value.

For direct projection, the mask of changed index~$j$ is
\begin{equation}
 x_T^{v+1}(j) = I(S_T^{v+1})_j\oplus I(A_T^v)_j.
 \label{eq:mask}
\end{equation}
A mask marks the stored bits that differ, so it reproduces the updated bits only where receiver storage holds the owner's baseline. That holds only on a \emph{repre\-sen\-ta\-tion-preserving} path, where the source projection $Q_T$ and the native loader preserve dtype and stored bits and insert no elements, and when no other write touches the destination. In the shared-scale example of \S~\ref{sec:residual}, multiplication and casting change stored bits, so overwrites carry the \mbox{converted values}.

\emph{Mixed} encoding therefore sends XOR entries $(d,x)$ for affine changes on representation-preserving paths whose destinations no other write touches, and overwrite entries $(d,o)$ for all other records, a choice fixed per mapping class and loader path. No uniform choice is both exact and compact: overwrites everywhere resend the unchanged bits of every changed value, and XOR everywhere corrupts values on paths that transform them. In both MoE checkpoints, XOR covers over \mbox{96\% of weight bytes}.

XOR entries also compress well. Small updates often leave high-order bits unchanged, so masks have many leading zeros, 14 in the mask \texttt{0x0003} of \S~\ref{sec:design} (Figure~\ref{fig:xorbits}, bottom). Run-length and gap coding shrink location bytes, and zstd~\cite{collet2021zstd} compresses \mbox{each payload}.

Yet XOR makes duplicate delivery unsafe. Applying a mask twice restores the old bits (Figure~\ref{fig:xorbits}, \mbox{bottom}), so receivers drop duplicates by identifier. A retry cannot tell which masks of a failed attempt took effect, so the retry \mbox{switches to overwrites}.

\section{Pipelined Delivery without a\texorpdfstring{\\*}{ }Cross-Cluster Collective}
\label{sec:protocol}

Following P3, delivery streams payloads without a cross-cluster collective (G4) and overlaps delta construction, application in synchronous RL, and rollout serving in asynchronous RL (G5). An NCCL broadcast would couple receiver membership to the training cluster and need a shared network route. Delivery~uses object storage or a relay tree, since some deployments share weights only via object storage~\cite{cursor2026composer} and others over direct \mbox{links~\cite{ye2026tensorhub}}.

\subsection{Object-storage and relay-tree delivery}
\label{sec:delivery}
\label{sec:codec-wire}

Payload bytes scale with emitted values, locations, and metadata, but transport topology and retries determine how often payloads cross the cluster boundary. Let $V_{\mathrm{payload}}$ be the bytes of unique compressed payloads before replication to receivers, and $V_{\mathrm{cross}}$ the bytes \mbox{crossing that boundary}.

\begin{figure}[!htbp]%
\centering
\includegraphics[width=\columnwidth]{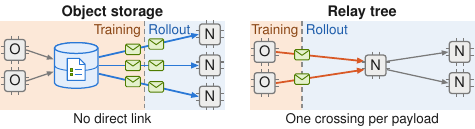}
\caption{Payloads move from owners~O to rollout nodes~N, with an envelope per boundary crossing.}
\label{fig:transports}
\end{figure}

\parhead{Object storage.}This transport trades boundary crossings for decoupling (Figure~\ref{fig:transports}, left): owners upload each payload once, list it in a manifest of expected payloads, and need no direct link to rollout nodes. Every rollout node then downloads each payload of the attempt across the cluster boundary, so~$V_{\mathrm{cross}}$ grows with the number of these nodes. A retry uploads the complete changed~set~again.

\parhead{Relay tree.}This transport streams each owner's payloads over direct links to one root in the rollout cluster (Figure~\ref{fig:transports}, right). The root forwards payloads through a balanced binary tree over local links, and~each node's receivers apply them. \mbox{Without retransmission}, payloads cross the cluster boundary once per attempt, so $V_{\mathrm{cross}}\approx V_{\mathrm{payload}}$.

\subsection{Overlapping refit stages}
\label{sec:pipeline}
\begin{figure}[!b]%
\centering
\includegraphics[width=\columnwidth]{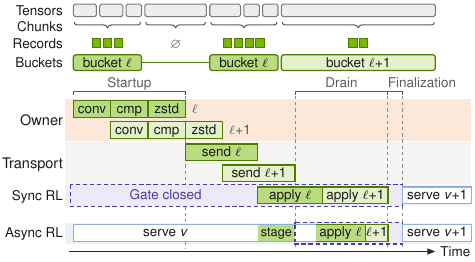}
\caption{Chunks group whole tensors, and buckets pack records (top). Pipelined refit stages of residual-task buckets (bottom).}
\label{fig:pipeline}
\end{figure}

Overlap matters because payload volume scales with the changed set, whereas delta construction cost scales with model size: construction scans training shards and converts required residual tensors. Building the whole update before sending it, as TensorHub, \mbox{AReaL-DTE}, and \mbox{SparseRL-Sync} all do~\cite{ye2026tensorhub,peng2026arealdte,hu2026sparserlsync}, leaves transport idle until construction ends. To overlap delta construction and delivery, owners group whole tensors into \emph{chunks} for \mbox{comparison} and pack their records into \emph{buckets}, each compressed into one payload (Figure~\ref{fig:pipeline}, top). Buckets skip unchanged chunks~($\varnothing$) and share metadata, compression, and transfer overhead \mbox{across chunks}.

Each stage hands a finished bucket to the next, so owners convert (conv), compare (cmp), and encode and compress (zstd) bucket $\ell{+}1$ while payload $\ell$ is sent, and receivers in synchronous RL apply $\ell$ as $\ell{+}1$ arrives (Figure~\ref{fig:pipeline}, bottom). Receivers that fall behind fill an apply queue of decoded payloads, and a full queue holds back delivery, capping receiver memory. Pipeline startup and \mbox{drain leave stages idle}.

A request gate keeps requests from mixing two policy versions and reopens only after \emph{finalization}: device synchronization, ACK, joint commit, and the in-place baseline update (Figure~\ref{fig:pipeline}, bottom). In synchronous RL, the coordinator closes it when the first attempt starts, so receivers can write weights during delivery. In asynchronous RL, rollout keeps serving version~$v$ while receivers stage the decoded delta in host memory. The gate closes only once the whole delta is staged, so after current requests finish, rollout pauses only for in-place application and finalization. Training takes its next optimizer step only after the transition ends, so retries recompute the delta from the same weights.

\section{Recoverable In-Place Refits}
\label{sec:apply}

After validating payloads and dropping duplicates, receivers update storage in place, which P4 makes safe (G6). A receiver-side baseline copy would double the stored weights: an overwrite retry repairs any partial update, and a joint commit keeps receiver storage and the source baseline in \emph{version alignment}, so each transition starts with every required receiver at the committed version. Placement stays with the native loader's rules for fused Q/K/V, routed experts, slicing, padding, and \mbox{tied weights (G1)}.\nowidowpar

\subsection{Applying deltas through the native loader}
\label{sec:apply-scratch}
\label{sec:apply-memory}

\begin{figure}[!htbp]%
\centering
\includegraphics[width=\columnwidth]{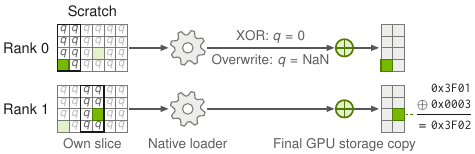}\\[5pt]
\includegraphics[width=\columnwidth]{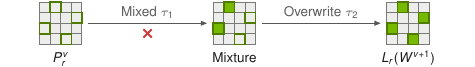}
\caption{Two of three ranks apply one item through scratch (top). Attempt~$\tau_1$ fails after writing two of four emitted destinations, and the overwrite retry~$\tau_2$ writes all four (bottom).}
\label{fig:applypath}\label{fig:retry}
\end{figure}

Reloading a patched checkpoint, as slime does~\cite{thudm2026slime}, rewrites every weight. To reuse the native loader without overwriting unchanged weights, receivers give it full canonical tensors and filter its writes. An \emph{item} is one owner's entries for one tensor. Each receiver scatters each item into a reusable host \emph{scratch} buffer of canonical shape and dtype, where positions without an entry are inactive and hold the placeholder~$q$ (Figure~\ref{fig:applypath}, top). The loader resolves names, keeps the rank's own slice, and transforms the scratch as in a dense refit. \sysname{} intercepts its final GPU storage copies, where \mbox{only active positions are updated}.

The placeholder marks positions to leave untouched. On the XOR path it is zero, since XOR with zero keeps stored bits. On the overwrite path it is a value absent from the item, such as NaN, and the non-placeholder positions at the first intercepted copy form an \emph{active mask} of the \mbox{values to write}.

\parhead{Limits of filtering.}Filtering is exact because, in supported loaders, every write of a dense refit passes through an intercepted copy and is a function of its own canonical value. A loader that mixes \mbox{values}, writes outside an intercepted copy, or skips a write that a dense refit performs breaks this, as do overlapping XOR writes. \sysname{} therefore checks loader paths in advance and fails any attempt whose copies break these conditions. Every model of \S~\ref{sec:evaluation} passes, and the checks guard against future \mbox{loader paths}.

\parhead{Validation and order.}Before application, receivers check that each payload belongs to the active attempt, verify its compressed-body digest, tensor names, shapes, dtypes, and item encodings, and drop duplicates by payload identifier. All receivers then apply payloads in one order that their identifiers declare, so a loader call with a cross-rank collective pairs the same payload on \mbox{every rank}.

\parhead{Receiver memory.}Receivers keep no baseline copy of the weights. In synchronous RL with the real updates of \S~\ref{sec:eval-crossstudy}, their host buffers take 15.6~GB per Qwen3-235B-A22B rollout node and would take \mbox{28.8~GB in asynchronous~RL}\@.

\subsection{Equivalence to a dense refit}
\label{sec:guarantee}

Let $\Delta^{v+1}$ encode $\mathcal{E}$, and let $\operatorname{apply}_r$ apply it to the storage~$P_r^v$ of required receiver~$r$.

\begin{proposition}[Dense-refit equivalence]\label{prop:bdt}
If a first attempt reaches its joint commit, replicas agree bitwise, the loader conditions of \S~\ref{sec:apply-scratch} hold, and the candidate weights $W^{v+1}$, committed baseline, source ownership, conversion rules, and loader settings stay fixed \mbox{until the commit, then}
\begin{equation}
  \operatorname{bits}\!\left(\operatorname{apply}_r(\Delta^{v+1}, P_r^v)\right) = \operatorname{bits}\!\left(L_r(W^{v+1})\right)\enspace\forall r,
  \label{eq:bdt}
\end{equation}
where $\operatorname{bits}$ reads all rank-local parameter and buffer storage but excludes the KV cache and other request state, which~each transition invalidates.
\end{proposition}

\parhead{Proof sketch.}Version alignment makes each XOR destination hold the owner's baseline bits, so deduplicated XOR entries reconstruct their updated values, and overwrite entries carry theirs. Under the loader conditions, filtering stores them as a dense refit would. Every changed destination has a record, and the others keep version-$v$ bits that match a dense refit because their canonical elements are unchanged, as Appendix~\ref{app:proof} \mbox{proves in full}.

\subsection{Overwrite recovery and joint commit}
\label{sec:commit}

Overwrite retries extend Proposition~\ref{prop:bdt} to transitions whose first attempt fails and leaves a mixture of old and updated values. Retrying with XOR masks would flip already updated bits back, and restoring periodic full checkpoints before replaying deltas, as PULSE does~\cite{miahi2026pulse}, stores and moves \mbox{full weights}. \sysname{} stops the failed attempt and waits until its writes can no longer take effect, then recomputes the complete changed set from the same candidate weights and unchanged $B^v$. Every record becomes an absolute overwrite, so the retry yields $L_r(W^{v+1})$ regardless of which earlier writes completed (Figure~\ref{fig:retry}, bottom). Receiver membership changes only between attempts, and~new or restarted receivers first restore the committed version from a peer that holds it \mbox{or by a dense refit}.

Owners compute each delta against the source baseline, so it fits only receivers at the same version. A single joint commit publishes the updated policy and binds it to its source baseline in a durable commit record. Compare-and-set updates this record only after every required receiver acknowledges, optionally after \emph{post-apply checks} of the written bits, so the policy never advances ahead of \mbox{any receiver}. Owners then update the task trackers and residual baseline in place before the transition ends, so the next delta uses the committed baseline. The commit record also makes failover safe: a~new coordinator reads it to decide whether to retry or finish the transition. These rules cover late, invalid, duplicate, and missing payloads, receivers killed mid-attempt, lost commit replies, and~coordinator failures, and a failure that persists across retries ends in \mbox{an operator abort (Appendix~\ref{app:api})}.

\section{Evaluation}
\label{sec:evaluation}

\noindent\textbf{RQ1.} How does \sysname{} compare with three other systems under real updates? \hfill\mbox{(\S~\ref{sec:eval-crossstudy})}\\
\textbf{RQ2.} How do latency and speedup change from 30B to 1T, and how large are payloads? \hfill\mbox{(\S~\ref{sec:eval-scale})}\\
\textbf{RQ3.} How do direct projection and mixed encoding change construction time and bytes? \hfill\mbox{(\S~\ref{sec:eval-ablation})}\\
\textbf{RQ4.} Does relay-tree transport dominate refit latency, and does delta construction overlap it? \hfill\mbox{(\S~\ref{sec:eval-floors})}\\
\textbf{RQ5.} Are refits bit-exact, and does training under receiver kills align with dense NCCL refits? \hfill\mbox{(\S~\ref{sec:eval-fidelity})}

\subsection{Experimental setup}
\label{sec:eval-setup}

\parhead{Testbed.}The latency testbed has eight training nodes with 32 NVIDIA GB300 GPUs in one AWS region and eight rollout nodes with 64 H100 GPUs in another. The object-storage transport is Amazon~S3\@. Each node's cross-cluster flow peaked at 5~Gbps. \mbox{The latency} runs of \S\S~\ref{sec:eval-scale} and~\ref{sec:eval-floors} use BF16 checkpoints of Nemotron-3 \mbox{Nano 30B}, \mbox{Super 120B}, and \mbox{Ultra 550B}, of a synthetic 1T model with twice the layers of \mbox{Ultra 550B}, and of Qwen3-235B-A22B, with sizes of \mbox{63.2--2{,}215~GB}, where 1~GB is $10^{9}$ bytes. Both transports use direct projection and mixed encoding with a 512~MiB bucket target and chunk targets of 64~MiB for object storage and 256~MiB for the relay tree, each tuned for the lowest refit latency. Latency and \S~\ref{sec:eval-crossstudy} runs use synchronous RL\@. In all runs, the native loader places every update (G1), and delivery goes over \mbox{object storage or the relay tree (G4)}.

\parhead{Element change rates.}Real change rates vary from step to step, so latency and delta construction runs fix the rate at 3\% or 5\%, stress cases that exceed the measured rates of Figure~\ref{fig:sparsity} and \S~\ref{sec:eval-crossstudy}. Changed source elements are chosen uniformly from each owner's shards, so almost every chunk holds changes. They are updated identically on all replicas with a relative amplitude of about $10^{-2}$.

\parhead{Timed window.}Refit latency spans every stage from change detection through transport and loader application to the end of finalization. It excludes post-apply checks and one-time setup such as model loading and initial \mbox{baseline construction}. Latency runs and those of \S~\ref{sec:eval-crossstudy} use five warm-up and ten timed refits per training-side parallel layout and replica count. The ten \sysname{} timings of each configuration span under 1~s, and we average them, then \mbox{weight every configuration equally}.

\parhead{Reference.}We use a transport-only full-checkpoint reference: of the dense approaches A and B in Table~\ref{tab:goals}, only full-checkpoint transfer meets G4, and the two clusters share no NCCL route over InfiniBand or Elastic Fabric Adapter (EFA)\@. Each training node uploads an eighth of the full \mbox{checkpoint} to object storage, and each rollout node then downloads all of it, the same path as \sysname{}'s \mbox{object-storage payloads}.

\subsection{Speedups of 7.0--99.5\texorpdfstring{$\times$}{\texttimes} over three systems}
\label{sec:eval-crossstudy}

Compared with ByteCheckpoint, \mbox{AReaL-DTE}, and PULSE, which take approaches B, C, and~D of Table~\ref{tab:goals}, \sysname{} is \mbox{70.8--99.5$\times$}, \mbox{7.0--7.3$\times$}, and \mbox{19.9--27.8$\times$} faster \mbox{(Table~\ref{tab:crossstudy})}. We time all four on the latency testbed under real updates from the GRPO recipe of \S~\ref{sec:sparsity}. In timed steps 6--15, these updates change about 2\% of elements per step on both Qwen3 models, more than in the first five steps of Figure~\ref{fig:sparsity}. These gaps match the challenges of \S~\ref{sec:intro}: ByteCheckpoint sends every weight, \mbox{AReaL-DTE} inverts AdamW to rebuild old weights and builds the whole update before sending it, and PULSE overlaps stages only when it replays missed versions. In the comparison runs, all three exchange tensors in the \mbox{Hugging Face} \mbox{checkpoint format}.

\emph{Takeaway:} The two other delta systems also send only changes, yet they are 7.0--27.8$\times$ slower. Sections~\ref{sec:eval-ablation} and~\ref{sec:eval-floors} measure the effect of \mbox{each design choice}.
\begin{table}[!t]
\caption{Mean object-storage refit latency (s) with real updates.}
\label{tab:crossstudy}
\centering
\footnotesize
\setlength{\tabcolsep}{2.5pt}
\begin{tabular}{@{}lrrrr@{}}
\toprule
Qwen3 & \sysname{} & ByteCheckpoint & AReaL-DTE & PULSE \\
\midrule
30B & 5.2 & 368.1 & 36.2 & 103.6 \\
235B & 28.5 & 2{,}836.9 & 208.7 & 793.4 \\
\bottomrule
\end{tabular}
\end{table}

\subsection{Refits from 30B to 1T are 12--40\texorpdfstring{$\times$}{\texttimes} faster}
\label{sec:eval-scale}

\begin{figure}[!htbp]%
\centering
\includegraphics[width=\columnwidth]{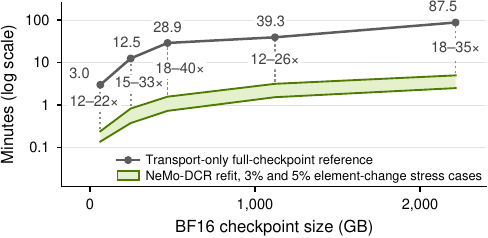}
\\[4pt]
\includegraphics[width=\columnwidth]{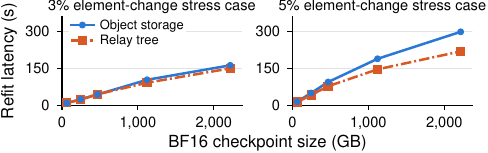}
\caption{Reference time in minutes and a band of \sysname{} latency, labeled with speedups (top). Mean latency per transport (bottom).}
\label{fig:fullckpt}
\end{figure}

Over object storage, which the reference also uses, \mbox{30B--1T} refits are \mbox{12--40$\times$} faster than the reference at 3\% and 5\% (Figure~\ref{fig:fullckpt}, top and bottom), while latency over both transports rises from \mbox{8.3--14.5~s} at 30B to \mbox{150--300~s} at 1T\@. The Seer rollout system cuts Kimi~K2 rollout from 12{,}202~s, 87\% of an RL step, to 8{,}267~s~\cite{qin2026seer}. With either rollout time, 1T refits of \mbox{150--300~s} in place of each step's weight update would make a synchronous step \mbox{1.4--1.5$\times$} faster than with the reference. At 120B, refits on both transports take \mbox{22.6--49.7~s} and are 15.1--33.2$\times$ faster (Table~\ref{tab:latency}). Their payload volume $V_{\mathrm{payload}}$ is 2.4\% and 3.8\% of the 247.2~GB checkpoint (G3), 21--24\% less than \mbox{the uncompressed changed values}.

\begin{table}[!t]%
\caption{Mean 120B refit latency and volume versus the transport-only full-checkpoint reference of \S~\ref{sec:eval-setup}.}
\label{tab:latency}
\centering
\footnotesize
\setlength{\tabcolsep}{2.2pt}
\begin{tabular}{@{}llrrr@{}}
\toprule
Method & Change rate & Latency $\downarrow$ & Volume $\downarrow$ & Speedup $\uparrow$ \\
\midrule
Object storage & 3\% & 24.5~s & 5.86~GB & 30.6$\times$ \\
Relay tree   & 3\% & 22.6~s & 5.82~GB & 33.2$\times$ \\
Object storage & 5\% & 49.7~s & 9.38~GB & 15.1$\times$ \\
Relay tree   & 5\% & 41.6~s & 9.35~GB & 18.0$\times$ \\
\midrule
Reference & N/A & 750~s & 247.2~GB & 1$\times$ \\
\bottomrule
\end{tabular}
\end{table}

At 5\%, the relay tree is 1.13$\times$ faster than object storage at 30B and 1.36$\times$ at 1T (Figure~\ref{fig:fullckpt}, bottom). The two stay close because every payload crosses the boundary into the tree root, as it does into each rollout node on object storage.

\emph{Takeaway:} Sending only compressed changes, 2.4--3.8\% of the checkpoint at 120B (G3, P2), makes refits fast on both \mbox{transports}.

\subsection{Projection saves time, XOR saves bytes}
\label{sec:eval-ablation}

\begin{table}[!htbp]%
\caption{Qwen3 delta construction time and payload volume. Each arrow names the only difference between the two variants it joins.}
\label{tab:bridgeablation}
\centering
{\footnotesize
\setlength{\tabcolsep}{2pt}
\begin{tabular}{@{}llrrrrr@{}}
\toprule
Qwen3 & Change rate & X (s) & P (s) & F (s) & X (GB) & P=F (GB) \\
\midrule
30B & 3\% & 2.96 & 2.77 & 3.20 & 1.49 & 2.44 \\
30B & 5\% & 3.23 & 2.99 & 3.29 & 2.29 & 3.80 \\
235B & 3\% & 20.23 & 18.38 & 19.87 & 11.37 & 18.42 \\
235B & 5\% & 23.07 & 20.42 & 21.96 & 17.58 & 29.36 \\
\bottomrule
\end{tabular}\par}
\vspace{4pt}
\includegraphics[width=\columnwidth]{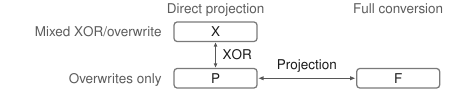}
\end{table}

For Qwen3, direct projection speeds up delta construction by \mbox{1.08--1.16$\times$} over full-tensor assembly and conversion, and mixed encoding cuts payload bytes by 38--40\% versus overwrites (P1 and P2, Table~\ref{tab:bridgeablation}). We~compare three variants in \mbox{standalone} delta \mbox{construction} runs of both stress cases on \mbox{Qwen3-30B-A3B} and \mbox{Qwen3-235B-A22B}. Variant X uses direct projection with mixed encoding, P the same with overwrites only, and F full conversion with overwrites only. Computing XOR masks from baseline bits adds 7--13\% to P's construction time (compare X with P in Table~\ref{tab:bridgeablation}).

A synthetic benchmark isolates value encoding with streams that hold the same changed values without locations or metadata. After level-1 zstd compression, XOR masks are 2.21$\times$, 2.18$\times$, and 1.72$\times$ smaller than overwrite values (P2) for noise of standard deviation $a=10^{-4}$, $10^{-3}$, and $10^{-2}$ added in FP32 to 20 million standard-normal BF16 values before rounding to BF16\@. Smaller updates leave more high-order bits unchanged and thus \mbox{zero in the mask}.

\emph{Takeaway:} Projection (P1) shortens construction by skipping full-tensor assembly. XOR masks (P2) cost a little construction time and cut bytes, a good trade because transport dominates refit \mbox{latency (\S~\ref{sec:eval-floors})}.

\subsection{Overlapped transport dominates latency}
\label{sec:eval-floors}

\begin{figure}[!htbp]%
\centering
\includegraphics[width=\columnwidth]{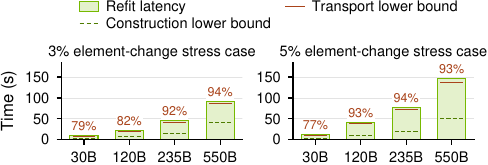}
\caption{Relay-tree refit latency and lower bounds, with the transport lower bound's share of latency.}
\label{fig:floors}
\end{figure}

Nsight Systems traces of 30B--550B relay-tree refits show that transport dominates refit latency and overlaps delta construction (P3, Figure~\ref{fig:floors}). The \emph{construction lower bound} is the longest per-owner union of delta-construction intervals, and the \emph{transport lower bound} the longest per-owner union of intervals from payload submission until the tree \mbox{root acknowledges queueing it}.

Across sizes and stress cases, the transport and construction lower bounds are \mbox{77--94\%} and \mbox{18--45\%} of refit latency and can sum to over 100\%, as the two overlap. Finalization, pipeline startup and drain, and transfer gaps leave 2.1--9.7~s outside the transport lower bound. Latency thus scales with payload bytes: in the stress cases of Table~\ref{tab:bridgeablation}, an overwrite retry carries \mbox{1.62--1.67$\times$} a first attempt's bytes, and in the receiver-kill runs of \S~\ref{sec:eval-fidelity}, retries take about 1.6$\times$ \mbox{as long as first attempts}.

\emph{Takeaway:} Because streaming (P3) hides construction behind transport, further gains need fewer bytes or more parallelism in \mbox{boundary crossings}.

\subsection{Training under kills\texorpdfstring{\hfil\penalty0\hfilneg\ }{ }aligns with dense NCCL refits}
\label{sec:eval-fidelity}

\begin{figure}[!htbp]%
\centering
\includegraphics[width=\columnwidth]{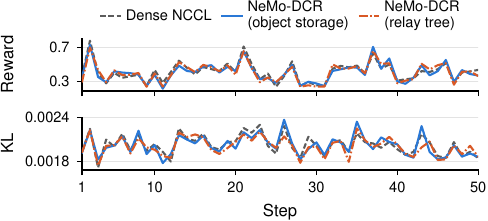}
\caption{Qwen3-30B-A3B reward and per-token KL, averaged over five runs. Mid-refit receiver kills occur only in \sysname{} runs.}
\label{fig:training}
\end{figure}

\sysname{} refits are bit-exact, and with receivers killed mid-refit, training on both transports aligns with the mean reward and KL trajectories of dense NCCL refits (P4, G2 and G6, Figure~\ref{fig:training}). In the latency runs, every \sysname{} refit passed its post-apply checks and matched a dense refit from the same candidate weights in every parameter and buffer bit. To exercise in-place application and overwrite retries on receivers that survive a kill, we train \mbox{Qwen3-30B-A3B} on agentic SWE tool calls for 50 asynchronous GRPO steps of 64 prompts with 8 responses each, in which rollout pauses only after the whole delta is staged. On a second testbed, training and rollout each use four H100 nodes, and its shared network lets us compare both \sysname{} transports with dense NCCL refits, approach~A of Table~\ref{tab:goals}. These updates change 1.2\% of elements per step on \mbox{average}.

Every five steps in each \sysname{} run, we pick a vLLM instance uniformly at random, kill it with its receivers mid-refit, and restart it five steps later by forking a receiver that holds the committed version. Each kill fails the in-flight attempt, and when it lands \mbox{during} application, the overwrite retry repairs the partial writes on the other receivers. Sampled responses vary, so we run each method five times, with independent kills in each \sysname{} run. KL is the estimated per-token KL divergence between rollout and training policies. Over 50 steps, mean reward is 0.41 and mean KL 0.0020 for dense NCCL refits and both \sysname{} transports, and every \sysname{} refit, retried or not, \mbox{passed its post-apply checks}.

\emph{Takeaway:} A mid-refit kill costs only an overwrite retry (P4), which ends with the bits of \mbox{a dense refit}.

\subsection{Limitations}
\label{sec:limitations}

\sysname{} spends construction work to cut transfer volume, so its gains depend on the change rate and the network: speedups shrink as more values change per step, and with 32 training GPUs, dense refits beat delta construction above about 200~Gbps across clusters. In asynchronous RL, staging holds one decoded delta per rollout node in host memory, about 4\% of the checkpoint for real Qwen3 updates. \sysname{} handles weights stored at any training precision. Supporting and validating FP8 or NVFP4 rollout is future work.

\section{Conclusion}
\label{sec:conclusion}

Agentic RL refits need not move full checkpoints: BF16 rounding leaves only 1--2\% of elements changed per step, and \sysname{} turns this sparsity into bit-exact delta refits. Describing changes in the canonical coordinates that training and serving share lets the native loader place them without model-specific logic. Bitwise encoding keeps changes exact, and XOR masks cut payload bytes by 38--40\% versus overwrites. Streaming hides construction behind transport and needs no cross-cluster collective. Retrying with overwrites makes in-place refits recoverable: with receivers killed midway, training aligns with \mbox{dense NCCL refits}. \sysname{} is 7.0--7.3$\times$ faster than \mbox{AReaL-DTE} with real updates, and 30B--1T refits are \mbox{12--40$\times$} faster than the reference%
, making delta refits practical for cross-cluster agentic RL at trillion-parameter scale. \sysname{}, with the \mbox{Megatron Bridge} and vLLM integration used in the experiments, is open source in NeMo RL~\cite{nemorl2026dcrcommit}.

\label{sec:main-end}
\phantomsection
\label{sec:references}
\pdfbookmark[1]{References}{bm:references}

  \begingroup\small\raggedright\setlength{\bibsep}{6pt plus 2pt minus 1pt}
  \bibliography{references}
  \endgroup

\appendix
\brokenpenalty=10000 %
\nowidows
\flushbottom%
\setlength{\textfloatsep}{6pt plus 8pt minus 2pt}\setlength{\intextsep}{6pt plus 8pt minus 2pt}\setlength{\floatsep}{6pt plus 8pt minus 2pt}%
\section*{Appendices}
\makeatletter
\@addtoreset{figure}{section}
\@addtoreset{table}{section}
\@addtoreset{algorithm}{section}
\@addtoreset{algocf}{section}
\makeatother
\renewcommand{\thefigure}{\thesection.\arabic{figure}}
\renewcommand{\thetable}{\thesection.\arabic{table}}
\renewcommand{\thealgorithm}{\thesection.\arabic{algorithm}}
\renewcommand{\thealgocf}{\thesection.\arabic{algocf}}
\providecommand{\theHalgocf}{}\renewcommand{\theHalgocf}{\thesection.\arabic{algocf}}

Appendix~\ref{app:codec} gives the direct-projection and XOR coverage and the format and volume of payloads. Appendix~\ref{app:proof} proves the dense-refit equivalence of Proposition~\ref{prop:bdt}, and Appendix~\ref{app:api} specifies the refit protocol and its failure handling. Appendices~\ref{app:integration}--\ref{app:estimates} give integration and measurement details, GRPO settings, the~evidence behind Table~\ref{tab:goals}, and the end-to-end estimates of \mbox{the main text}.

\Needspace{4\baselineskip}
\section{Coverage and the\texorpdfstring{\hfil\penalty0\hfilneg\ }{ }Payload~Format}
\label{app:codec}

For \S\S~\ref{sec:bdt}, \ref{sec:codec}, and~\ref{sec:delivery}, this appendix details the coverage of direct projection and XOR, the payload format, and how payload volume and traffic relate to the element change rate and the number of rollout nodes.

\begin{table}[!b]
\centering
\footnotesize
\caption{\mbox{Nemotron-3-Ultra-550B-A55B} tensors and byte shares by weight type and task.}
\label{tab:affinecoverage}
\renewcommand{\arraystretch}{1.05}
\setlength{\tabcolsep}{3pt}
\begin{tabular}{@{}>{\raggedright\arraybackslash}m{0.60\columnwidth}rr@{}}
\toprule
\textbf{Weight type} & \textbf{Tensors} & \textbf{Bytes} \\
\midrule
\multicolumn{3}{@{}l}{\emph{Affine}} \\
Routed expert up/down & 50{,}176 & 93.86\% \\
Other affine weights & 662 & 3.13\% \\
\addlinespace
\emph{Residual} & & 3.01\% \\
Attention Q/K/V & 39 & \\
Mamba input projection and~convolution & 144 & \\
Embedding and output head & 2 & \\
\bottomrule
\end{tabular}
\end{table}

\begin{table}[!t]%
\centering
\footnotesize
\caption{\mbox{Qwen3-30B-A3B} tensors and byte shares by weight type and task, affine or residual.}
\label{tab:qwencoverage}
\setlength{\tabcolsep}{3pt}
\begin{tabular}{@{}llrr@{}}
\toprule
\textbf{Weight type} & \textbf{Task} & \textbf{Tensors} & \textbf{Bytes} \\
\midrule
Expert gate/up/down & affine & 18{,}432 & 94.95\% \\
Attention output projection & affine & 48 & 1.32\% \\
Router and normalization & affine & 241 & 0.04\% \\
Attention Q/K/V & residual & 144 & 1.65\% \\
Embedding and output head & residual & 2 & 2.04\% \\
\bottomrule
\end{tabular}
\end{table}

\paragraph{Direct-projection coverage.} Table~\ref{tab:affinecoverage} shows that the affine mappings cover all but 185 canonical tensors of the Nemotron-3-Ultra-550B-A55B checkpoint, which holds 1{,}121~GB\@. Besides routed experts, its affine weights include shared experts, routers, latent-MoE projections, attention output projections, normalization, and the output projections and state parameters of Mamba. Its conversion has no custom export hook, which would postprocess converted tensors and disable direct projection. Routed~experts use per-expert Auto mappings, which infer their parallel type from the Megatron module and meet the conditions of Appendix~\ref{app:sourceconditions} for direct projection. Of~these 185 residual tensors, 183 use unsupported mapping classes: three Q/K/V tensors per attention layer, including the multi-token-prediction layer, and three input projection and convolution tensors per Mamba layer. The~other two are the embedding and output head, whose Megatron Bridge conversion drops vocabulary padding. Table~\ref{tab:qwencoverage} shows that 18{,}721 of the 18{,}867 canonical tensors of \mbox{Qwen3-30B-A3B} are affine, and its residual tensors are attention Q/K/V and, owing to padding, the embedding and output head. Each of these residual conversions only moves elements, so neither checkpoint exercises the shared \mbox{scale factors of \S~\ref{sec:residual}}.

\begin{figure}[!htbp]%
\centering
\includegraphics[width=\columnwidth]{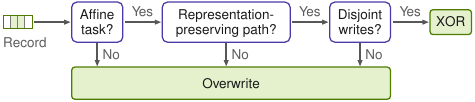}\\[5pt]
\includegraphics[width=\columnwidth]{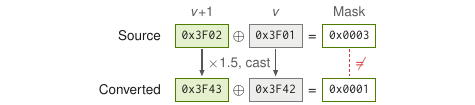}
\caption{How a record gets its encoding in mixed mode (top). A source XOR mask cannot update scaled and cast values (bottom).}
\label{fig:xorrule}\label{fig:convmask}
\end{figure}

\paragraph{XOR rule.} Mixed encoding gives all records of an item the same encoding. On the default BF16 paths of Appendix~\ref{app:loaderconditions}, each owner's part of an affine canonical tensor with changed elements contributes one item. The item is an XOR item if its path is representation-preserving with disjoint writes, and an overwrite item otherwise. Each required residual tensor emits an overwrite item when its converted bits change. Figure~\ref{fig:xorrule}, top, summarizes the rule, and the bottom shows why scaled and cast values need overwrites. A conversion scales the values of the \S~\ref{sec:design} example by 1.5 and casts them to BF16\@. Their source mask is \texttt{0x0003}, but that of the converted values \texttt{0x3F42} and \texttt{0x3F43} is \texttt{0x0001}, so the source mask cannot update \mbox{the converted bits}.

\begin{figure}[!htbp]%
\centering
\includegraphics[width=\columnwidth]{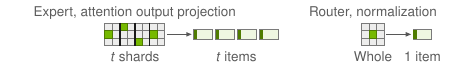}
\caption{XOR items of Qwen3-30B-A3B with every shard changed, drawn for parallel degree~$t=4$.}
\label{fig:items}
\end{figure}

\paragraph{XOR coverage.} For the Qwen3-30B-A3B checkpoint, XOR covers 96.3\% of weight bytes at every training-side tensor-parallel and expert-tensor-parallel degree, because a higher degree only splits the same weights into more shards. With every tensor and shard changed, XOR covers 18{,}721 of 18{,}867 items at degree one, 74{,}161 of 74{,}307 at degree four, and 148{,}081 of 148{,}227 at degree eight. Each of the 18{,}432 expert tensors and 48 attention output projections splits into $t$ shards when both degrees are $t$, while the 48 router and 193 normalization tensors stay whole, giving \mbox{$18{,}480t+241$ XOR items (Figure~\ref{fig:items})}.

For the Nemotron-3-Ultra-550B-A55B checkpoint of Table~\ref{tab:affinecoverage}, every affine item takes the XOR path except those of the 48 Mamba \texttt{A\_log} tensors, which~the~native loader stores as their negated exponential. These tensors are tiny, so the XOR byte share rounds to the 97.0\% affine share. With~every tensor and shard changed, XOR covers 404{,}150 of the 404{,}719 items at tensor-parallel and expert-tensor-parallel degree eight. At expert-tensor-parallel degree $t_{\mathrm{e}}$, each of the 50{,}176 routed expert up/down tensors splits into $t_{\mathrm{e}}$ shards, and at tensor-parallel degree $t$, 352 of the 662 other affine tensors split into $t$ shards, while the remaining 310 stay whole. These $50{,}176t_{\mathrm{e}}+352t+310$ affine items are all XOR items except the $48t$ \texttt{A\_log} items, and each of \mbox{the 185 residual tensors} \mbox{adds one overwrite item}.\nowidowpar

\begin{figure}[!b]%
\centering
\includegraphics{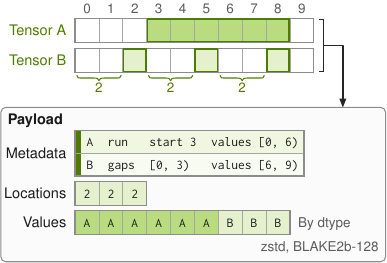}\\[5pt]
\includegraphics[width=\columnwidth]{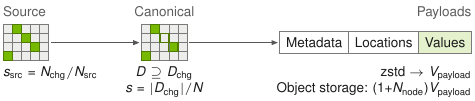}
\caption{The payload format with a contiguous and a scattered tensor (top). The quantities behind the payload volume (bottom).}
\label{fig:payload}\label{fig:volume}
\end{figure}

\paragraph{Payload format.} Each payload carries attempt, owner, and payload identifiers, flat location and value buffers, and~per-item metadata: tensor name, shape, dtype, encoding, location type, and offsets into both buffers. Contiguous locations use a start index, and the number of values gives the run length. Scattered locations use gaps in the smallest suitable integer type. A BLAKE2b-128 digest~\cite{aumasson2013blake2} covers each compressed payload body. Figure~\ref{fig:payload}, top, illustrates the payload format with two tensors. The changed elements of tensor A form one contiguous run, so its metadata row stores a start index instead of location offsets and adds nothing to the location buffer. Tensor B's scattered positions are stored as gaps, which decode from $-1$ by adding each gap plus one, so gaps 2, 2, 2 yield positions 2, 5, 8. Both rows index the shared value buffer, whose values have \mbox{the width of their tensor's dtype}.

\paragraph{Change density.} To relate payload bytes to the element change rate (Figure~\ref{fig:volume}, bottom), let $N=|\Omega|$ be the element count of the full canonical index set~$\Omega$. Because ownership is disjoint, $N_{\mathrm{src}}=\sum_T|J_T|$ counts unique source elements, $N_{\mathrm{chg}}=\sum_T|C_T|$ counts unique changes, and their ratio $s_{\mathrm{src}}=N_{\mathrm{chg}}/N_{\mathrm{src}}$ is the element change rate. The emitted destinations form \mbox{$D=\bigcup_T\pi_T(C_T)\cup\bigcup_{u\in\mathcal{U}_{\mathrm{req}}}D_u$} over affine tasks~$T$. The destinations with changed canonical bits form \mbox{$D_{\mathrm{chg}}=\{d\in\Omega\mid I(W^{v+1})_d\ne I(W^v)_d\}$}, and complete change detection makes it a subset of $D$. The~canonical change density is $s=|D_{\mathrm{chg}}|/N$, which equals the element change rate when source and canonical elements correspond one to one along representation-preserving paths. Other mappings can add unchanged destinations to~$D$, such as the one drawn outlined in \mbox{Figure~\ref{fig:volume}, bottom}.

\paragraph{Payload bytes.} Before zstd, the payloads of one attempt together carry metadata, run-length- and gap-coded locations, and~one value per record. Let~$H$~be the canonical checkpoint size $\sum_{d\in\Omega}b_d$, where $b_d$ is the width of element $d$ in bytes. With one record per emitted destination, the value bytes total $\sum_{d\in D}b_d$. At uniform width $b_d=H/N$, this sum is $|D|H/N$, or $sH$ when $D=D_{\mathrm{chg}}$. A destination with several overwrite-encoded records adds another $b_d$ bytes for each record beyond the first. More scattered changes need more location bytes, and more items need more metadata. After zstd compresses each payload, the attempt's payloads total $V_{\mathrm{payload}}$. On the object-storage transport, each attempt moves about $(1+N_{\mathrm{node}})V_{\mathrm{payload}}$ of upload and download traffic when $N_{\mathrm{node}}$ rollout nodes download every payload. Only the downloads cross the \mbox{cluster boundary, so $V_{\mathrm{cross}}\approx N_{\mathrm{node}}V_{\mathrm{payload}}$}.\nowidowpar

\FloatBarrier%
\section{Dense-Refit Equivalence}
\label{app:proof}

\begin{figure}[!t]%
\centering
\includegraphics[width=\columnwidth]{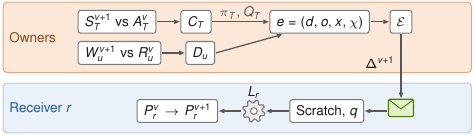}
\caption{Notation of Table~\ref{tab:notation} on the refit path.}
\label{fig:notationmap}
\vspace{8pt}\end{figure}

Appendices~\ref{app:sourceconditions} and~\ref{app:loaderconditions} state the conditions of Proposition~\ref{prop:bdt}. Appendix~\ref{app:equivproof} proves the proposition and extends it to overwrite retries of failed attempts. Table~\ref{tab:notation} and Figure~\ref{fig:notationmap} summarize the notation \mbox{used throughout the paper}.

\begin{table}[!htbp]%
\caption{Main notation. Other symbols are defined in the sections where they are used.}
\label{tab:notation}
\centering
\footnotesize
\setlength{\tabcolsep}{3pt}
\renewcommand{\arraystretch}{1.05}
\begin{tabular}{@{}>{\raggedright\arraybackslash}p{0.2\columnwidth}>{\raggedright\arraybackslash}p{0.61\columnwidth}l@{}}
\toprule
\textbf{Symbol} & \textbf{Meaning} & \textbf{Where} \\
\midrule
$W^v$ & Canonical parameter and buffer tensors at version $v$ & \ref{sec:boundary} \\
$L_r$, $P_r^v$ & Receiver $r$'s native loader and receiver storage $L_r(W^v)$ & \ref{sec:boundary} \\
$B^v$ & Committed source baseline: task trackers and residual baseline & \ref{sec:detect} \\
$S_T^{v+1}$, $A_T^v$ & Candidate shard and tracker of task $T$ & \ref{sec:detect} \\
$J_T$, $C_T$ & Owned and changed source indices of task $T$ & \ref{sec:detect} \\
$I(X)_j$ & Stored bit pattern of $X$ at index $j$ & \ref{sec:detect} \\
$d=(n,k)$ & Canonical destination: tensor name and flat index & \ref{sec:projection} \\
$\pi_T$, $Q_T$ & Index mapping and source projection of an affine task & \ref{sec:projection} \\
$\mathcal{U}_{\mathrm{req}}$ & Required tensor set of an attempt & \ref{sec:residual} \\
$W_u^{v+1}$, $R_u^v$, $D_u$ & Converted tensor, residual baseline, and changed destinations of tensor $u$ & \ref{sec:residual} \\
$\mathcal{E}$ & Complete changed set of delta records & \ref{sec:residual} \\
$(d,x)$, $(d,o)$ & XOR entry and overwrite entry & \ref{sec:codec} \\
$V_{\mathrm{payload}}$ & Compressed payload bytes before replication & \ref{sec:delivery} \\
$V_{\mathrm{cross}}$ & Bytes crossing the cluster boundary & \ref{sec:delivery} \\
$q$ & Placeholder at inactive scratch positions & \ref{sec:apply-scratch} \\
$\Delta^{v+1}$ & Encoding of $\mathcal{E}$ applied at each receiver & \ref{sec:guarantee} \\
$\operatorname{apply}_r$, $\operatorname{bits}$ & Delta application at receiver $r$ and stored bits of rank-local parameters and buffers & \ref{sec:guarantee} \\
$K^v$ & Commit record naming version $v$ and its baseline & \ref{app:sourceconditions} \\
$e$ & Delta record: destination, overwrite value, XOR mask, and path tag & \ref{app:sourceconditions} \\
$\mathcal{T}_{\mathrm{aff}}$, $\mathcal{T}_{\mathrm{res}}$, $\mathcal{U}$ & Affine tasks, residual tasks, and residual-baseline tensors & \ref{app:sourceconditions} \\
$g$, $\mathcal{M}$ & Candidate-baseline identifier and integration setup & \ref{app:sourceconditions} \\
$\rho_r(e)$ & Destination byte ranges written on receiver $r$ for record $e$ & \ref{app:loaderconditions} \\
$\tau$ & Attempt identifier, returned by \texttt{prepare} & \ref{app:api} \\
$m$ & Encoding mode: mixed or overwrite & \ref{app:api} \\
\bottomrule
\end{tabular}
\end{table}

\subsection{Source conditions and version state}
\label{app:sourceconditions}

Proposition~\ref{prop:bdt} needs these source and version-state conditions: each unique source element has one owner and one conversion task, owners emit a record for every changed canonical destination, the committed baseline stays fixed until the joint commit, and every required receiver starts each transition at the version that the commit record names.

\begin{figure}[!htbp]%
\centering
\includegraphics[width=\columnwidth]{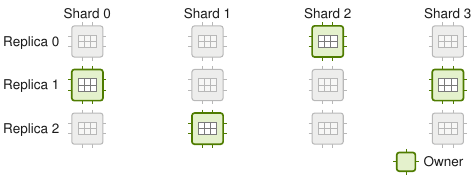}\\[5pt]
\includegraphics[width=\columnwidth]{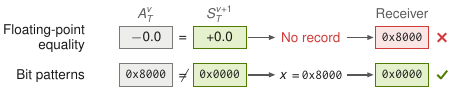}
\caption{One owner per shard across data-parallel replicas (top). Comparing bit patterns catches a zero that changes sign (bottom).}
\label{fig:ownership}\label{fig:bitdiff}
\end{figure}

\paragraph{Ownership.} Each shard has one owner among its replicas. The parallel layout identifies replica groups, which include the data-, context-, and expert-data-parallel replicas and the tensor-parallel copies of replicated parameters. Every training step keeps the copies within each replica group bitwise identical, so one copy of each shard is enough: a stable name hash selects it as the owner and spreads the owners across the replicas, and the other copies emit no records (Figure~\ref{fig:ownership}, top). Tied weights likewise have one owner and one task.

\begin{figure}[!t]%
\centering
\includegraphics[width=\columnwidth]{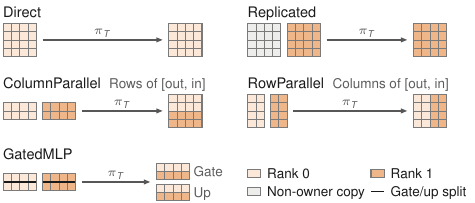}
\caption{Affine mapping classes for two ranks.}
\label{fig:mappings}
\end{figure}

\paragraph{Affine tasks.} In Megatron Bridge, the affine mapping classes determine canonical coordinates given local indices, shard geometry, parallel ranks, and global expert indices~\cite{nvidia2026megatronbridge}. Figure~\ref{fig:mappings} draws them for two tensor-parallel ranks. Direct maps one tensor one to one, ColumnParallel shards hold rows and RowParallel shards hold columns of a weight of shape [out,~in], and each GatedMLP shard holds gate rows above up rows. Only the owner's copy of a Replicated tensor is projected, and each rank's shard maps through~$\pi_T$ to fixed canonical coordinates. \mbox{For direct projection}, the Megatron Bridge integration supports only these affine classes and those Auto mappings that permute no dimensions and act as ColumnParallel, \mbox{RowParallel}, or \mbox{Replicated} mappings. All of them use Megatron Bridge metadata for shard axes, offsets, and gate/up splitting. A custom export hook postprocesses converted tensors, \mbox{so it disables direct projection}.

\paragraph{Residual tasks and change flags.} Tied embeddings, the output head that depends on them, and any source tensor that also feeds a residual conversion always use residual tasks. A residual task can combine several owners' inputs, so its change flag must combine theirs. Each owner sets a task's flag when its shard of the task changes and contributes zero for tasks it does not hold, drawn gray in Figure~\ref{fig:flags}, top. The \texttt{all\_reduce\_flags} collective of Appendix~\ref{app:api} combines the flags by task identifier over all owners. Before assembly, the flags and conversion dependencies, drawn as arrows, fix the attempt's required tensor set $\mathcal{U}_{\mathrm{req}}\subseteq\mathcal{U}$, where $\mathcal{U}$ indexes the residual-baseline tensors. In the figure, $u_4$ depends only on the unflagged~$T_3$ and keeps its \mbox{committed value}. A custom export hook leaves conversion dependencies unknown, so any flagged task \mbox{selects every residual tensor}.

\paragraph{Residual assembly.} Residual tasks whose outputs form one canonical tensor are converted together when that tensor is in $\mathcal{U}_{\mathrm{req}}$. For example, per-expert tasks are converted together when the checkpoint stores all experts as one stacked tensor. Each owner's part of a residual tensor $u$ carries the tensor name, shape, dtype, and canonical indices it covers. The parts must jointly cover every element of~$u$, and overlapping parts must agree \mbox{bit-for-bit}. In the assembly of Figure~\ref{fig:assembly}, bottom, one changed element in expert~$E_1$ of a stacked expert tensor~$u$ makes every owner convert its part. The owner selected by a hash~$h$ of the canonical name~$n_u$ assembles $W_u^{v+1}$, and the comparison with $R_u^v$ puts only the changed element into~$D_u$. Figure~\ref{fig:residual}, top, contrasts \mbox{this path with direct projection}.\nowidowpar

\begin{figure}[!b]%
\centering
\includegraphics[width=\columnwidth]{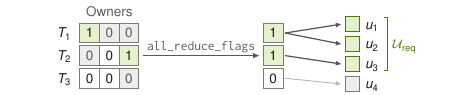}\\[5pt]
\includegraphics[width=\columnwidth]{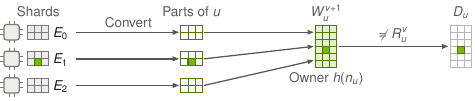}
\caption{Change flags of $T_1$--$T_3$ select required residual tensors (top). Residual assembly of a stacked expert tensor~$u$ (bottom).}
\label{fig:flags}\label{fig:assembly}
\end{figure}

\paragraph{Change detection.} Change detection compares stored bit patterns, because floating-point equality misses a zero that changes sign. Equality treats $+0.0$ and $-0.0$ as equal although their BF16 patterns \texttt{0x0000} and \texttt{0x8000} differ. A zero that changes sign would then get no record, and receivers would keep the old sign bit, as Figure~\ref{fig:bitdiff}, bottom, shows for \texttt{0x8000}. Comparing bit patterns finds the change, and the XOR mask \texttt{0x8000} flips the receiver's sign bit to give \texttt{0x0000}, \mbox{the value a dense refit writes}.\nowidowpar

\begin{figure}[!t]
\centering
\includegraphics[width=\columnwidth]{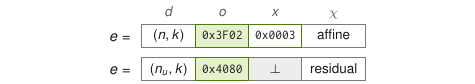}
\caption{Affine and residual delta records.}
\label{fig:records}
\end{figure}

\paragraph{Records.} The affine and residual paths both emit records $e=(d,o,x,\chi)$ in $\mathcal{E}$ (Figure~\ref{fig:records}). Each record holds a destination~$d$, an overwrite value~$o$, an XOR mask~$x$, which is $\bot$ for residual records, and a path tag $\chi\in\{\mathsf{affine},\mathsf{residual}\}$. The example's affine record carries the overwrite value \texttt{0x3F02} and the XOR mask \texttt{0x0003}. For a residual record, the codec encodes only the overwrite value, such as \mbox{\texttt{0x4080} in Figure~\ref{fig:records}}.

\newcommand{\tabadmission}{%
\begin{table*}[!t]
\centering
\footnotesize
\caption{Operations, integration rules, and assumptions for the default BF16 source and loader paths, where views include transpositions.}
\label{tab:loaderadmission}
\renewcommand{\arraystretch}{1.14}
\setlength{\tabcolsep}{4pt}
\begin{tabular}{@{}>{\raggedright\arraybackslash}p{0.12\textwidth}>{\raggedright\arraybackslash}p{0.26\textwidth}>{\raggedright\arraybackslash}p{0.28\textwidth}>{\raggedright\arraybackslash}p{0.28\textwidth}@{}}
\toprule
\textbf{Path} & \textbf{Operations} & \textbf{Integration rule} & \textbf{Path-specific assumptions} \\
\midrule
\multicolumn{4}{@{}l}{\emph{Source}} \\
Affine & Direct, Replicated, ColumnParallel, RowParallel, GatedMLP, and qualifying~Auto & Supported mapping classes; no custom export hook & Fixed deterministic \mbox{Megatron Bridge} mappings; stored bits preserved for XOR~entries \\
\addlinespace
Residual & Other mapping classes, stacking, padding, tied weights, adapters, custom postprocessing, and quantization scales & Residual conversion; forced overwrite & Complete dependency metadata, or every residual tensor selected when~unknown \\
\midrule
\multicolumn{4}{@{}l}{\emph{Loader}} \\
XOR & Identity, views, slices, splits, fusions & Copy input shares scratch storage; same dtype; no within-item overlap & Stored bits preserved; no overlap across items \\
\addlinespace
Overwrite & Identity, views, slices, splits, fusions, casts, padding, pointwise transforms such as~$-\exp(\cdot)$ & Copy input is floating-point; active values distinct from the placeholder; active mask remains valid; overlapping writes of different entries agree bitwise & Placeholders preserved until each view's first copy; each output depends only on its own input \\
\addlinespace
Native skip & Rank-local skips and tied weights & Skip recorded by \texttt{prewarm}; reported items match intercepted copies & Accurate reports; interception of every storage write \\
\bottomrule
\end{tabular}
\end{table*}}

\begin{figure}[!htbp]%
\centering
\includegraphics[width=\columnwidth]{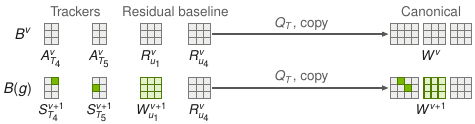}
\caption{The committed and candidate baselines and the canonical tensors that they reconstruct.}
\label{fig:baseline}
\end{figure}

\paragraph{Version state.} For affine tasks $\mathcal{T}_{\mathrm{aff}}$ and residual tasks $\mathcal{T}_{\mathrm{res}}$, the committed source baseline is
\[ B^v=\{A_T^v\}_{T\in\mathcal{T}_{\mathrm{aff}}\cup\mathcal{T}_{\mathrm{res}}}\cup\{R_u^v\}_{u\in\mathcal{U}}. \]
The baseline $B^v$ reconstructs $W^v$: the source projections $Q_T$ map the affine-task trackers into canonical coordinates, and the residual-baseline tensors supply the rest. The commit record $K^v=(v,g_v)$ names version~$v$ and the identifier $g_v$ of its baseline $B^v$. With $W^{v+1}$ and $B^v$ fixed, the \texttt{prepare} call of Appendix~\ref{app:api} binds a fresh candidate-baseline identifier $g$ to the attempt. The logical candidate baseline $B(g)$ replaces every tracker $A_T^v$ by its candidate shard $S_T^{v+1}$ and the residual baseline $R_u^v$ of every $u\in\mathcal{U}_{\mathrm{req}}$ by its converted tensor $W_u^{v+1}$, and keeps $R_u^v$ for every other~$u$. $B(g)$ reconstructs $W^{v+1}$ whenever every changed residual tensor is in $\mathcal{U}_{\mathrm{req}}$, as Figure~\ref{fig:baseline} shows. There, $B(g)$ holds the candidate shards of the affine tasks $T_4$ and~$T_5$, with changed elements in dark green. It also holds, in light green, the converted $W_{u_1}^{v+1}$ of the required~$u_1$ of Figure~\ref{fig:flags}, top, \mbox{while $u_4$ keeps $R_{u_4}^v$}.

\begin{figure}[!htbp]%
\centering
\includegraphics[width=\columnwidth]{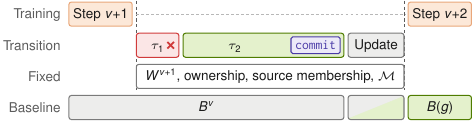}\\[5pt]
\includegraphics[width=\columnwidth]{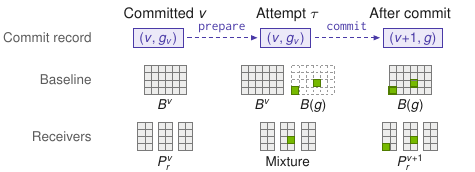}
\caption{State that stays fixed across the attempts of a transition (top). Version state at three points of a transition (bottom).}
\label{fig:fixedstate}\label{fig:versions}
\end{figure}

\paragraph{Fixed state.} The candidate weights $W^{v+1}$, source ownership and membership, and integration setup $\mathcal{M}$ remain fixed across the attempts of the transition, and $B^v$ remains fixed until commit. The integration setup $\mathcal{M}$ contains the conversion rules with their mapping and dependency metadata, the transport adapter of Appendix~\ref{app:implementation}, and the native loaders with their settings, XOR/overwrite/skip classifications, and the validated placeholder schemes of Appendix~\ref{app:loaderconditions}. To~keep $W^{v+1}$ fixed, training takes its next optimizer step only after the transition ends (Figure~\ref{fig:fixedstate}, top). After the commit, the baseline update, shaded gray to green, turns $B^v$ into~$B(g)$. Source membership is the set of training ranks, and changing it requires redistributing or reinitializing \mbox{the baseline before another transition}.

\paragraph{Version alignment.} The protocol preserves \emph{version alignment} between the source baseline and receiver storage: when a transition starts, every required receiver holds $P_r^v$ for the version~$v$ that the commit record names (Figure~\ref{fig:versions}, bottom). During an attempt, the commit record and $B^v$ stay fixed, the candidate baseline $B(g)$, drawn dashed, exists only logically, and receivers may hold a mixture of old and updated values. Receivers~obtain $P_r^v$ from the committed transition to version~$v$, from a fork that copies a receiver holding it, or from a dense refit, and receiver storage changes only through refits and forks. To keep this alignment, a commit requires completion acknowledgments from every required receiver, and~a new or restarted receiver joins only after restoring the committed version (Appendix~\ref{app:failure}).

\subsection{Loader conditions and properties}
\label{app:loaderconditions}

Proposition~\ref{prop:bdt} also needs the loader conditions of \S~\ref{sec:apply-scratch} and two loader properties, detailed here.

\paragraph{Loader conditions.} Exact delta application requires intercepting every storage write: every item that the loader reports as loaded must have an intercepted copy, and every other item, such as a routed expert on another rank, must match a native skip recorded by \texttt{prewarm}, a loader dry run. The loader must transform each canonical value independently and deterministically. XOR~byte ranges must be disjoint from all other writes, and overlapping overwrites must yield the same bits as a dense refit. \sysname{} rejects \mbox{unsupported} loader paths in advance and fails an attempt if any copy breaks these conditions. For each source and loader path, Table~\ref{tab:loaderadmission} lists the operations it allows, the rules that \sysname{}'s \mbox{Megatron Bridge} and vLLM integration enforces, and the path-specific assumptions \mbox{that Proposition~\ref{prop:bdt} requires}.\nowidowpar

\tabadmission%

\begin{figure}[!htbp]%
\centering
\includegraphics[width=\columnwidth]{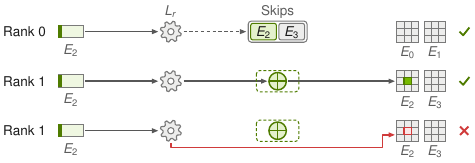}
\caption{A native skip, an intercepted write, and a write past the dispatch hook for a routed expert~$E_2$.}
\label{fig:skips}
\end{figure}

\paragraph{Native skips.} The \texttt{prewarm} call runs the loader with tensor metadata to find what each receiver skips, such as tied weights or routed experts on other ranks. All other writes must pass through the dispatch hook, which intercepts the loader's final GPU storage copies. In Figure~\ref{fig:skips}, every receiver gets the routed-expert item~$E_2$. The loader on rank~0, which holds $E_0$ and~$E_1$, skips $E_2$ as \texttt{prewarm} recorded, and on rank~1 it writes $E_2$ through an XOR copy~$\oplus$ in the dashed dispatch hook. In the bottom lane, rank~1 writes $E_2$ past the hook, so the reported item has no intercepted copy \mbox{and the attempt fails}.

\begin{figure}[!b]%
\centering
\includegraphics[width=\columnwidth]{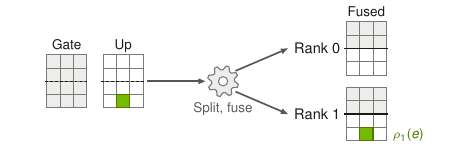}
\caption{A fused default loader path on two ranks.}
\label{fig:loaderpath}
\end{figure}

\paragraph{Write overlap.} Let $\rho_r(e)$ be the destination byte ranges that receiver~$r$ writes when applying the entry encoded from record~$e$. The two overlap conditions above, disjoint XOR byte ranges and dense-refit bits under overlapping overwrites, apply to these ranges across entries, items, payloads, and loader calls. An XOR copy fails the attempt if its input no longer shares scratch storage, its input and destination dtypes differ, or its destination bytes overlap another copy in the item. Runtime checks for XOR overlap reset at each item and miss writes from other items, so advance validation ensures that XOR writes do not overlap them. A loader change requires another \texttt{prewarm} call and revalidation of these conditions and the loader properties. On the fused path of Figure~\ref{fig:loaderpath}, the native loader splits the canonical gate and up projections across two tensor-parallel ranks at the dashed lines and fuses them per rank at the solid lines. Each output element has one canonical source, and a changed up-projection element has one destination range in \mbox{$\rho_1(e)$ on rank~1}.

\begin{figure}[!t]%
\centering
\includegraphics[width=\columnwidth]{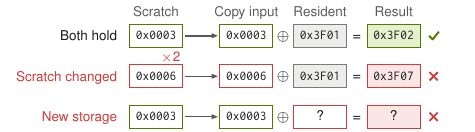}\\[5pt]
\includegraphics[width=\columnwidth]{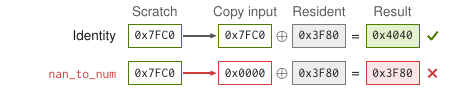}
\caption{An XOR copy needs both loader properties (top). \texttt{nan\_to\_num} erases an XOR mask with a NaN bit pattern (bottom).}
\label{fig:loaderprops}\label{fig:xormask}
\end{figure}

\paragraph{Loader properties.} Beyond the loader conditions, the integration must establish two loader properties: scratch bits remain unchanged before an XOR copy, and model parameters and buffers keep their existing storage. When both hold, the intercepted copy of the example XORs the mask into the resident version-$v$ bits and yields \texttt{0x3F02} (Figure~\ref{fig:loaderprops}, top). An in-place doubling of scratch would alter the mask, and new parameter storage would lack the resident bits, so~either gives wrong bits. The integration therefore validates both properties for \mbox{the configured loader in advance}.

\paragraph{XOR copies.} On an XOR path, the loader copies XOR masks, so to stay representation-preserving it must insert no elements and must preserve every bit pattern that a mask can contain at its dtype width, including NaN, infinity, subnormal, and signed-zero patterns. In Figure~\ref{fig:xormask}, bottom, the mask between the resident 1.0 (\texttt{0x3F80}) and the updated 3.0 (\texttt{0x4040}) is \texttt{0x7FC0}, a BF16 NaN pattern. An identity path keeps the mask and yields 3.0, whereas \texttt{nan\_to\_num} turns the mask into \texttt{0x0000} and leaves 1.0 unchanged. Operations such as \texttt{nan\_to\_num}, clipping, casts, and \mbox{padding use overwrite only}.

\paragraph{Overwrite copies.} On an overwrite path, the first intercepted copy to each destination view forms its active mask from the non-placeholder positions of its input, and only active positions are written. The default overwrite paths require a floating-point copy input and use NaN as the placeholder, so every NaN bit pattern marks an inactive position. Other schemes compare bit patterns with the placeholder. If an active value matches the placeholder, such as an active NaN under the default scheme, the receiver scatters the item again with the next validated scheme of its loader path before the loader runs, so the attempt needs no retry. An item that has an active value matching the placeholder of every such scheme fails the attempt. Active~values are cast to the destination dtype. A transform that erases the placeholder before the active mask is formed, such as \texttt{nan\_to\_num}, requires another placeholder scheme, which the integration validates \mbox{as described below}.

\begin{figure}[!htb]%
\centering
\includegraphics[width=\columnwidth]{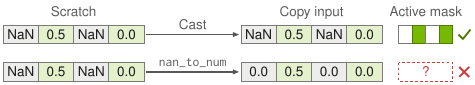}\\[5pt]
\includegraphics[width=\columnwidth]{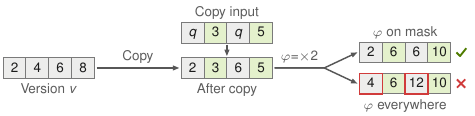}
\caption{A cast keeps the NaN placeholder, and \texttt{nan\_to\_num} erases it (top). A later pointwise transform~$\varphi$ reuses the active mask of the first copy (bottom).}
\label{fig:placeholder}\label{fig:maskreuse}
\end{figure}

\paragraph{Placeholder schemes.} A placeholder scheme picks the placeholder~$q$ that marks inactive positions on an overwrite path. To validate one, the integration traces placeholders and active values through \texttt{nan\_to\_num}, clipping, casts, and the other operations of the loader path. Outputs must be deterministic and depend only on their corresponding inputs across supported shapes, dtypes, and value ranges, under fixed loader settings and cast behavior. The transformed placeholder must remain distinct from all active values until the active mask is formed. The integration setup $\mathcal{M}$ lists each loader path's validated schemes in order. If no scheme passes for a loader path, \texttt{prepare} rejects it. In Figure~\ref{fig:placeholder}, top, with placeholders in gray and active values in green, a cast keeps the NaN placeholders, so the non-NaN positions of the copy input form the active mask. But \texttt{nan\_to\_num} maps the placeholders to 0.0, which is also an active value, so the mask cannot be formed and the path \mbox{needs another validated placeholder scheme}.\nowidowpar

\paragraph{Mask reuse.} Within an item, later intercepted writes of a pointwise loader transform may reuse the active mask of their destination view when they target the same storage, offset, shape, and strides. Such a transform acts in place: it reads the view after the first copy and writes a function of each stored value back through another intercepted copy, as the doubling~$\varphi$ of Figure~\ref{fig:maskreuse}, bottom, does. Advance validation fixes which intercepted writes of a path form a mask and which reuse one, and the hook drops its masks after each item. Only active positions are written, so the values that the transform computes elsewhere are discarded. The ordered writes of that transform must preserve the active mask and produce the same final bits as a dense refit (Figure~\ref{fig:maskreuse}, bottom). Applied everywhere, the doubling would transform the inactive positions, outlined in red, a second time, because their version-$v$ values are already transformed. A transform that instead updates the view with an in-place operation and no intercepted copy bypasses the dispatch hook, so advance validation rejects \mbox{its path}.

\subsection{Equivalence proof}
\label{app:equivproof}

\begin{figure}[!htbp]%
\centering
\includegraphics[width=\columnwidth]{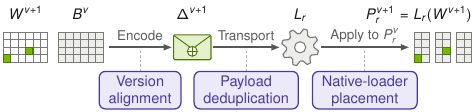}
\caption{The chain of Proposition~\ref{prop:bdt} from $W^{v+1}$ and $B^v$ to the storage of three receivers, with the guarantee used at each arrow.}
\label{fig:transition}
\end{figure}

The full proof behind the sketch of \S~\ref{sec:guarantee} combines complete change detection and exact XOR and \mbox{overwrite} entries with the three guarantees of Figure~\ref{fig:transition}: version alignment between the source baseline and receiver storage~$P_r^v$, payload deduplication at ingestion, and native-loader \mbox{placement at the receiver}.

\paragraph{Full statement of Proposition~\ref{prop:bdt}.} Proposition~\ref{prop:bdt} names its main assumptions, and the full statement adds the rest. For a first attempt that passes every check up to and including the \texttt{flush} call of Appendix~\ref{app:api}, receiver storage satisfies~\eqref{eq:bdt} under the following assumptions (Table~\ref{tab:assumptions}), and it still does at the joint commit, because \texttt{commit} writes no receiver storage. The candidate weights $W^{v+1}$, baseline $B^v$, source ownership and membership, and integration setup $\mathcal{M}$ remain fixed until commit, and $\mathcal{M}$ includes the conversion rules and loader settings. Each required receiver initially holds $P_r^v$, and receiver membership stays fixed during the attempt. Replicated source copies agree bit-for-bit, and the owners cover each unique source element exactly once. Every required residual tensor is fully assembled, with overlapping parts agreeing bit-for-bit. Both projected and overwrite values are computed deterministically, and every XOR-encoded record uses a representation-preserving path. The residual dependency metadata is complete, or unknown dependencies select every residual tensor, so that $W_u^{v+1}$ equals $R_u^v$ bit-for-bit for every $u\notin\mathcal{U}_{\mathrm{req}}$. The value written to each destination byte range depends only on its canonical element, if it has one. The~loader classification covers every final GPU storage copy, and the loader conditions of \S~\ref{sec:apply-scratch}, the path-specific assumptions of Table~\ref{tab:loaderadmission}, and the loader properties of \mbox{Appendix~\ref{app:loaderconditions} hold}.

\begin{table}[!b]%
\caption{Assumptions of Proposition~\ref{prop:bdt}, how each is established, and the figures that illustrate it.}
\label{tab:assumptions}
\centering
\footnotesize
\setlength{\tabcolsep}{2.8pt}
\renewcommand{\arraystretch}{1.05}
\begin{tabular}{@{}>{\raggedright\arraybackslash}p{0.37\columnwidth}>{\raggedright\arraybackslash}p{0.43\columnwidth}>{\raggedright\arraybackslash}p{0.15\columnwidth}@{}}
\toprule
\textbf{Assumption} & \textbf{Established by} & \textbf{Figure} \\
\midrule
$W^{v+1}$, $B^v$, source ownership and membership, and $\mathcal{M}$ fixed until commit & The next optimizer step waits for the transition; owners update $B^v$ only after commit; training ranks, ownership, and $\mathcal{M}$ change only between transitions & \ref{fig:fixedstate} (top) \\
Receivers start from $P_r^v$; receiver membership fixed per attempt & Commit only after all required receivers acknowledge; receivers join after a fork or dense refit and only between attempts & \ref{fig:versions} (bottom), \ref{fig:membership} (bottom) \\
Copies agree; one owner per source element & Identical replica updates; name-hash ownership & \ref{fig:ownership} (top) \\
Residual tensors fully assembled and consistent & Checks of \texttt{build\_complete} in Table~\ref{tab:api} & \ref{fig:assembly} (bottom) \\
Deterministic values; XOR only on representation-preserving paths & Fixed conversion rules; encoding decided per mapping class and loader path & \ref{fig:xorrule} (top) \\
Dependencies complete, or every residual tensor selected & Change flags with dependency metadata & \ref{fig:flags} (top) \\
Each written value depends only on its canonical element & Supported loader paths & \ref{fig:loaderpath} \\
Every final GPU storage copy classified; loader conditions, Table~\ref{tab:loaderadmission} assumptions, and properties hold & \texttt{prewarm} skip discovery, copy interception, and advance validation & \ref{fig:skips}; both halves of \ref{fig:loaderprops} and \ref{fig:placeholder} \\
\bottomrule
\end{tabular}
\vspace{4pt}\end{table}

\newcommand{\figxorinv}{%
\begin{figure}[!t]%
\centering
\includegraphics[width=\columnwidth]{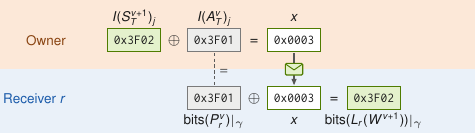}\\[5pt]
\includegraphics[width=\columnwidth]{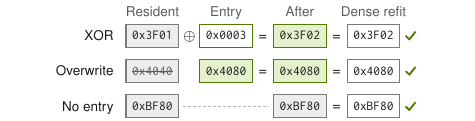}
\caption{The XOR invariant for the example element (top). The three kinds of receiver byte ranges in the proof (bottom).}
\label{fig:xorinv}\label{fig:proofcases}
\end{figure}}

\paragraph{XOR invariant.} For each XOR-encoded record $e$ from affine task $T$ and source index $j$, the \emph{XOR invariant} states that every range $\gamma\in\rho_r(e)$ holds the matching source-baseline bits until its XOR copy:
\[
\operatorname{bits}(P_r^v)\big|_{\gamma}=I(A_T^v)_j,
\qquad e=(\pi_T(j),o,x,\mathsf{affine}),
\]
where each range $\gamma$ spans one stored value. Version alignment ensures that the receiver holds these bits: its storage is at version $v$, the baseline version used to compute the XOR mask. Under the proposition's assumptions, the projection $Q_T$ and loader preserve dtype and stored bits, and no other entry writes to that range, so the invariant holds. In Figure~\ref{fig:xorinv}, top, the dashed equality marks the invariant for the example. Because it holds, the owner sends only the mask~$x$ across the cluster boundary and no \mbox{baseline values}.

\paragraph{Proof.}
Under the assumptions of the proposition, owners compare every unique source element and required residual tensor bitwise, affine outputs depend only on their source elements, and residual tensors outside $\mathcal{U}_{\mathrm{req}}$ keep their committed bits. Hence every changed canonical destination has a record. Mixed encoding uses XOR only where no other write overlaps the destination, so an XOR-encoded record is its destination's only record. Any other emitted destination has one record or several overwrite-encoded records with identical value bits: conversion is deterministic and overlapping parts agree, which the \texttt{build\_complete} check of \mbox{Appendix~\ref{app:api}} verifies within each bucket. Payload deduplication, together with the arrival check and queue draining of \texttt{flush} in Appendix~\ref{app:delivery}, ensures that each required receiver applies every XOR and overwrite entry \mbox{of the attempt exactly once}.

For an XOR-encoded record $e$ from affine task $T$ and source index $j$, the mask is $x=I(S_T^{v+1})_j\mathbin{\oplus}I(A_T^v)_j$, and scratch carries the item's XOR masks at changed positions and zeros elsewhere. For every $\gamma\in\rho_r(e)$, the invariant and the representation-preserving path make $\operatorname{bits}(P_r^v)|_{\gamma}\mathbin{\oplus}x$ equal $\operatorname{bits}(L_r(W^{v+1}))|_{\gamma}$, the bits that a dense refit writes.

For overwrite entries, a validated placeholder scheme establishes the active mask, and the overwrite-copy rules of Appendix~\ref{app:loaderconditions} require every pointwise loader transform's ordered writes to preserve it. The complete overwrite sequence thus writes only at active positions, and as every written value depends only on its canonical element, if any, it writes \mbox{$\operatorname{bits}(L_r(W^{v+1}))$ there}.

By native-loader placement, every byte range of an emitted destination is written and holds its target bits, and under the loader conditions and properties every other byte range keeps its version-$v$ bits (Figure~\ref{fig:proofcases}, bottom). Because every changed canonical destination has a record, each such range depends on no canonical element, as padding does, or only on an unchanged one, so its bits already equal $\operatorname{bits}(L_r(W^{v+1}))$. XOR-written, overwrite-written, and unwritten byte ranges together cover all storage that \mbox{$\operatorname{bits}$ reads, which proves~\eqref{eq:bdt}}.

\figxorinv%

\paragraph{Retries.}
Overwrite retries extend Proposition~\ref{prop:bdt} to transitions whose first attempt fails. Every attempt \mbox{recomputes} $\mathcal{E}$ deterministically from the same $W^{v+1}$, $B^v$, ownership, and~$\mathcal{M}$, so all attempts of a transition emit the same destinations. After a failed attempt's \texttt{mark\_obsolete} and \texttt{stop\_and\_wait} calls of Appendix~\ref{app:api} complete, a retry that passes every check up to and including \texttt{flush} has written every byte range of these destinations as an absolute overwrite. If every write of the earlier attempts went through the dispatch hook, those attempts changed only these ranges. Such a retry then turns any mixture they left into $\operatorname{bits}(L_r(W^{v+1}))$ before commit (Figure~\ref{fig:retry}, bottom), so retries are safe to repeat. A write that bypasses the hook can reach bytes outside these ranges. Its attempt fails (Figure~\ref{fig:skips}, bottom lane), as does every retry with the same receiver, since the loader repeats the write. \mbox{After an} operator abort as in Appendix~\ref{app:failure}, a fork or \mbox{a dense refit restores that receiver}.

\section{Refit Protocol}
\label{app:api}

\begin{table*}[!t]
\caption{Protocol operations.}
\label{tab:api}
\centering
\footnotesize
\setlength{\tabcolsep}{3pt}
\renewcommand{\arraystretch}{1.0}
\begin{tabular}{@{}>{\raggedright\arraybackslash}p{0.43\textwidth}>{\raggedright\arraybackslash}p{0.54\textwidth}@{}}
\toprule
\textbf{Operation} & \textbf{Effect} \\
\midrule
\multicolumn{2}{@{}l}{\emph{Source and coordinator}} \\
\texttt{refit}$(W^{v+1},B^v,K^v,\mathcal{M})$ & Run a transition with overwrite recovery \\
\texttt{prepare}$(v{+}1,\mathcal{M},g,m)\to\tau$ & Validate $\mathcal{M}$, then bind $g$, $m$, and required receivers to a fresh attempt \\
\texttt{build\_complete}$(W^{v+1},B^v,\mathcal{M},g)\to\{\mathcal{E}_\ell\}_\ell$ & Stream changed-set buckets under $g$ and check each for completeness and clashing canonical values \\
\texttt{bitdiff}$(X,Y)$ & Compare equal-width stored values and return changed indices \\
\texttt{all\_reduce\_flags}$(\{(T,f_T)\})\to\mathcal{U}_{\mathrm{req}}$ & Reduce change flags within the training cluster to find the required residual tensors \\
\texttt{convert}$(T,u)$ & Produce task $T$'s part of canonical tensor $u$ \\
\texttt{encode}$(\mathcal{E}_\ell,m)\to(p_\ell,\sigma_\ell)$ & Encode bucket $\mathcal{E}_\ell$ as compressed XOR or overwrite entries \\
\texttt{send}$(\tau,p_\ell)$ & Send identified payload $p_\ell$ under $\tau$ \\
\texttt{wait\_for\_requests}$(\tau,v)$ & Close the request gate if open, then drain version-$v$ requests \\
\texttt{open\_apply}$(\tau)$ & Let receivers apply staged and queued payloads \\
\texttt{flush}$(\tau)\to(\mathit{ack},\sigma)$ & Check arrivals, drain apply queues, synchronize devices, and collect acknowledgments with any configured post-apply checks \\
\texttt{commit}$(\tau)\to c$ & Conditionally replace $K^v$ with $(v{+}1,g)$ after \texttt{flush} \\
\texttt{read\_commit}$(\tau)\to K$ & Return the commit record once no commit of $\tau$ can change it \\
\texttt{mark\_obsolete}$(\tau)$ & Invalidate $\tau$ for later payloads, \texttt{open\_apply}, and \texttt{commit} \\
\texttt{stop\_and\_wait}$(\tau)\to\mathit{ack}$ & Stop delivery, then wait for writes and acknowledgments \\
\texttt{cleanup}$(\tau)$ & Release the payloads and any remaining delta records \\
\texttt{reopen\_gate}$(\tau)$ & Invalidate KV caches, then reopen the request gate \\
\midrule
\multicolumn{2}{@{}l}{\emph{Receiver and transport endpoints}} \\
\texttt{prewarm}$(\mathcal{M})\to\sigma$ & Reserve scratch and discover native skips \\
\texttt{ingest}$(\tau,\mathit{id},p)$ & Validate and deduplicate, then queue decoded payloads, or stage them if the attempt stages \\
\texttt{apply}$(p)$ & Scatter into scratch, then call the native loader \\
\bottomrule
\end{tabular}
\end{table*}

For both transports of \S~\ref{sec:protocol} and the joint commit and overwrite recovery of \S~\ref{sec:commit}, this appendix specifies the protocol operations and their failure handling.

The active coordinator runs one transition at a time, and each transition takes one or more attempts. Each attempt has a fresh attempt identifier $\tau$ and candidate-baseline identifier $g$, under which the complete changed set $\mathcal{E}$ is streamed. Appendix~\ref{app:sourceconditions} defines the committed baseline $B^v$, its commit record $K^v$, the candidate baselines, and the state \mbox{that remains fixed across retries}.

In the signatures of Table~\ref{tab:api}, $\mathcal{M}$ is the fixed integration setup of Appendix~\ref{app:sourceconditions}, and operations that take an attempt identifier~$\tau$ act only on that attempt. Mode $m$ is mixed or overwrite, and the latter mode encodes every record as an absolute overwrite. The symbols $p$, $\mathit{id}$, $\mathit{ack}$, and $\sigma$ denote a payload, its identifier, an acknowledgment, and a $\mathsf{success}$ or $\mathsf{failed}$ status, and $c$ is a commit outcome as defined in Appendix~\ref{app:failure}. Task $T$'s change flag is $f_T$. Inside \texttt{build\_complete}, \texttt{bitdiff} detects changes, owners directly project affine changes into canonical coordinates, \texttt{all\_reduce\_flags} fixes the required residual tensors $\mathcal{U}_{\mathrm{req}}$, and \texttt{convert} produces each residual task's part of these tensors. A second \texttt{bitdiff} then compares each assembled tensor \mbox{with its residual baseline}.

An attempt's payloads, end markers, control calls, and gate operations carry the attempt identifier $\tau$ that its \texttt{prepare} returned. Once a later \texttt{prepare} has replaced that attempt, receiver and transport endpoints ignore them, except that they acknowledge a \texttt{stop\_and\_wait} for the replaced attempt at once, so the call never waits on those endpoints.

\paragraph{Failure model.} Owners, receivers, transports, and coordinators may crash or lose messages, but they do not act maliciously, and channels are trusted or authenticated. Table~\ref{tab:failures} lists how the protocol handles each failure. Section~\ref{sec:eval-fidelity} exercises receiver kills, and the other rows follow from the protocol. A control plane, assumed available and clock-synchronized with coordinators, holds the durable commit record, accepts \texttt{prepare} and the end of each transition only from the active coordinator, and raises a $\mathsf{conflict}$ exception for any later \texttt{prepare} or \texttt{commit} \mbox{of a replaced coordinator}.

\begin{table}[!b]%
\caption{Failures and their handling.}
\label{tab:failures}
\centering
\footnotesize
\setlength{\tabcolsep}{3pt}
\renewcommand{\arraystretch}{1.0}
\begin{tabular}{@{}>{\raggedright\arraybackslash}p{0.27\columnwidth}>{\raggedright\arraybackslash}p{0.50\columnwidth}>{\raggedright\arraybackslash}p{0.17\columnwidth}@{}}
\toprule
\textbf{Failure} & \textbf{Handling} & \textbf{Figure} \\
\midrule
Rejected \texttt{prepare} & Before admission, ends the transition; after admission, triggers overwrite recovery & \ref{fig:modes} \\
Payload of an obsolete attempt & Dropped at reception & \ref{fig:reception} (top) \\
Invalid payload & The attempt fails before the payload is applied & \ref{fig:reception} (top) \\
Duplicate payload & Ignored if its bytes match; otherwise the attempt fails & \ref{fig:reception} (top) \\
Missing payload or end marker & The arrival check or the delivery deadline fails the attempt, and an overwrite retry follows & \ref{fig:commitseq} (top) \\
Receiver killed mid-attempt & Retired with its serving instance by \texttt{stop\_and\_wait}; the instance rejoins after a fork or dense refit, and the retry updates the others & \ref{fig:retry} (bottom), \ref{fig:membership} (bottom) \\
Lost commit reply & After the deadline, \texttt{read\_commit} decides the outcome & \ref{fig:readcommit} (bottom) \\
Coordinator failure & Before admission, takeover starts in mixed mode; after admission, it drains the attempt, calls \texttt{read\_commit}, then retries or, if the old commit took effect, finishes the transition & \ref{fig:failover} (top) \\
Owner loses its state & Operator abort; the owner reinitializes its baseline & \ref{fig:operator} (bottom) \\
Other persistent failure & Operator abort; a closed gate stays closed until a fork or dense refit & \ref{fig:operator} (bottom) \\
\bottomrule
\end{tabular}
\end{table}

\subsection{Admission and delivery}
\label{app:delivery}

Each~attempt runs \texttt{prepare}, then streams, checks, applies, and flushes its payloads as Algorithm~\ref{alg:schedule} specifies for synchronous and asynchronous RL (\S~\ref{sec:pipeline}). The coordinator spawns the algorithm's three tasks and runs its remaining lines. The attempt stages payloads if it starts in asynchronous RL with the request gate open. Each receiver loop ends once it has handled every owner's end marker and the payloads that the marker counts. An owner whose encoding fails still finishes \texttt{build\_complete} without sending, so the training ranks leave their collectives together. A failed attempt, including one that misses its delivery deadline, is retried in overwrite \mbox{mode (Appendix~\ref{app:failure})}.\nowidowpar

\begin{algorithm2e}[!b]%
\caption{Concurrent schedule of one attempt.}
\label{alg:schedule}
\DontPrintSemicolon
\Input{Prepared $\tau$, $g$, $v$, $m$, $W^{v+1}$, $B^v$, $\mathcal{M}$, \mbox{bounded queues}}
\Output{$(\mathit{ack},\mathsf{success})$, $(\bot,\mathsf{failed})$, or a raised $\mathsf{failed}$}
\Spawn{owners}{
  \ForAll{buckets $\mathcal{E}_\ell$ streamed by \texttt{build\_complete}}{
    $(p_\ell,\sigma_\ell)\gets\texttt{encode}(\mathcal{E}_\ell,m)$\;
    \lIf{$\sigma_\ell\ne\mathsf{success}$}{finish \texttt{build\_complete} without sending, then \Raise $\mathsf{failed}$}
    \texttt{send}$(\tau,p_\ell)$\;
  }
  send an end marker with the payload count under $\tau$\;
}
\Spawn{receiver ingestion}{
  \ForAll{arriving $(\mathit{id},p)$}{
    \texttt{ingest}$(\tau,\mathit{id},p)$\;
  }
}
\Spawn{receiver application}{
  \WaitUntil the coordinator calls \texttt{open\_apply}$(\tau)$\;
  \lIf{the attempt stages}{\KwSty{spawn} queueing of staged payloads in declared order}
  \ForAll{$p$ from the apply queue in declared order}{
    \texttt{apply}$(p)$\;
  }
}
\If{the attempt stages}{
  \Await all sends\;
  \WaitUntil every node has staged the delivered payloads\;
  \Raise $\mathsf{failed}$ if any expected payload is missing\;
}
\texttt{wait\_for\_requests}$(\tau,v)$\;
\texttt{open\_apply}$(\tau)$\;
\Await all sends if the attempt does not stage\;
\Return{\texttt{flush}$(\tau)$}\;
\end{algorithm2e}

\begin{figure}[!ht]%
\centering
\includegraphics[width=\columnwidth]{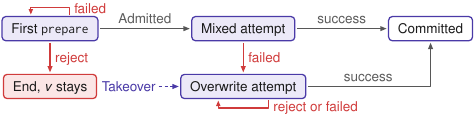}
\caption{Admission and attempt modes.}
\label{fig:modes}
\end{figure}

\paragraph{Admission.} The \texttt{prepare} call checks the metadata and loader classifications in $\mathcal{M}$, then binds $g$, mode $m$, and the set of required receivers to the fresh attempt. The first successful \texttt{prepare} admits the transition. A \texttt{prepare} whose checks do not pass raises $\mathsf{reject}$, a \texttt{prepare} of a replaced coordinator raises $\mathsf{conflict}$, and any other error raises $\mathsf{failed}$. Before admission, a rejection ends the transition and leaves $B^v$ as the next \mbox{transition's baseline}. After a $\mathsf{failed}$ exception before admission, the coordinator retries the first \texttt{prepare} in mixed mode. After admission, receivers may already hold partial writes, which a repeated XOR mask would flip back, so every $\mathsf{reject}$ or $\mathsf{failed}$ exception triggers overwrite recovery. Overwrite attempts also encode XOR-path records as overwrites. Every XOR path preserves all bit patterns and thus the NaN placeholder, so the default scheme is valid for these records. A transition starts in mixed mode, or~in overwrite mode when a coordinator takes over an admitted transition after failover. Retries stop at commit success, a conflict, or a rejection before admission, and an operator abort also ends them. Figure~\ref{fig:modes} draws these modes with takeover after admission, \mbox{leaving out conflicts and operator aborts}.

\paragraph{Transports.} Both transports follow Algorithm~\ref{alg:schedule} but carry payloads differently. On object storage, owners upload each payload as a multipart object. Payloads and the manifest remain in storage until \texttt{cleanup}, which follows \texttt{commit} or a completed \texttt{stop\_and\_wait}. The relay tree streams payloads over ZeroMQ DEALER/ROUTER sockets. Each node keeps a bounded queue for each owner's payloads and forwards an owner's next payload to a child only when the child's queue for that owner has room for it. Each attempt builds its tree over the nodes of its required receivers at \texttt{prepare}, so after a node fails, even the root, the retry sends payloads over a new tree without that node.\nowidowpar

\paragraph{Streaming and staging.} In synchronous RL, in which rollout waits for the refit, the coordinator follows \texttt{prepare} by calling \texttt{wait\_for\_requests} and then \texttt{open\_apply}, so receivers decode payloads into the bounded apply queue and apply them while later payloads arrive. In~asynchronous RL, an attempt that starts with the request gate open instead stages: each node's receivers decode validated payloads into one shared staged delta in host memory while rollout keeps serving version~$v$. Once every node has decoded every expected payload into its staged delta, the coordinator closes the gate, drains \mbox{version-$v$} requests, and calls \texttt{open\_apply}. Staged payloads then feed the apply queue in declared order. A retry that starts with the gate closed streams as in synchronous RL, because rollout \mbox{no longer serves} requests.\nowidowpar

\begin{figure}[!htbp]%
\centering
\includegraphics[width=\columnwidth]{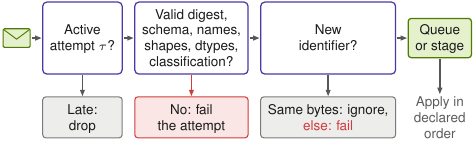}\\[5pt]
\includegraphics[width=\columnwidth]{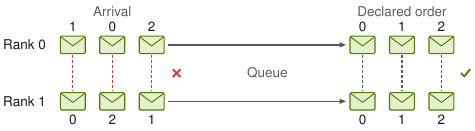}
\caption{Reception checks before a payload is queued or staged (top). Applying in declared order pairs the same payload in each cross-rank collective (bottom).}
\label{fig:reception}\label{fig:order}
\end{figure}

\paragraph{Reception checks.} When \texttt{prepare} starts an attempt, receivers learn the active attempt's version, identifier, and owners. Before queueing or staging a payload, they check that it belongs to the active attempt, so they drop late payloads from obsolete attempts. Receivers then check each payload's compressed-body digest, tensor names, shapes, dtypes, and schema, and match each item's encoding against the loader classification in~$\mathcal{M}$. If any of these checks does not pass, the attempt fails before that payload is applied. Receivers key each payload by its attempt, owner, and \mbox{payload} identifiers. They apply it at most once and ignore identical duplicates. Reusing~an identifier with \mbox{different} bytes fails the attempt. Figure~\ref{fig:reception}, top, summarizes \mbox{these checks and their outcomes}.

\paragraph{Application order.} After these reception checks, all receivers apply payloads in the single order that the payload identifiers declare for the attempt, so that a loader call with a cross-rank collective pairs the same payload on every rank, whatever the arrival order. In Figure~\ref{fig:order}, bottom, payloads numbered by their position in declared order arrive at two receiver ranks in different orders, so applying on arrival would pair different payloads in a collective, drawn dashed. Each owner numbers its payloads consecutively from zero and ends its stream with an end marker that carries its payload count, which is zero if it emits no records. An owner without records forms no bucket and sends only its end marker, so the declared order skips that owner from the start. The declared order takes payloads by number, then by owner, and skips an owner once its end marker shows that it has no payload with that number. Each owner sends its end marker once it has compared all its shards and any residual tensors it assembles, since only then is its payload count fixed. If an owner sends its last payload before it finishes comparing, later payloads in declared order wait until its end marker shows that the owner has no more, and the bounded queues hold the payloads that arrive \mbox{meanwhile}. Each owner sends in number order, and receivers consume each owner's payloads in that order, so the next payload in declared order is always its owner's first unconsumed payload. Within a payload, receivers may reorder only \mbox{independent} loader operations that have no cross-rank collectives or \mbox{cross-payload side effects}.

\paragraph{Flow control.} The bounded queues never stall application in declared order. The required receivers of a node share the bounded apply queue, and a payload leaves it only after all of them have applied it. An assembly group collects decoded payloads before they join the queue. The queue and its assembly group keep a slot for the earliest payload in declared order that some receiver of the node has yet to apply, and application can take that payload from either. Once only that slot is free, ingestion queues nothing else but still reads end markers and drops duplicates, so payloads later in declared order cannot fill the queue and stall application. Ingestion reads a payload's identifiers before decoding it. Queue slots free one payload at a time, not one group of payloads at a time. Relay-tree hops queue each owner's stream separately, so holding back one stream does not block the others. On object storage, receivers fetch the next payload in declared order themselves. Owners are coupled through \texttt{all\_reduce\_flags} and residual assembly, so one blocked owner would stall them all. Hence \texttt{send} never waits for a held-back stream: it keeps the unsent payload in host memory, and the owner keeps constructing later buckets while the held-back \mbox{stream waits for queue space}.

\paragraph{Clashes.} Invalid payloads, clashing canonical values, or clashing destination writes stop application and fail the attempt, whether it is the first attempt or a retry. Clashes across owners or items arise only when an assumption of \mbox{Proposition}~\ref{prop:bdt} fails. Each~source element has one owner and replicas agree, so canonical values cannot clash. Each written value depends only on its canonical element, so overlapping overwrite writes carry equal bits. Advance validation keeps the XOR writes of different items disjoint. The runtime checks therefore need no state across items: \texttt{build\_complete} checks each bucket, and receivers verify the writes of each item they apply.

\paragraph{Flush.} The \texttt{flush} call closes the attempt to new sends, waits for delivery and relay-tree forwarding, and checks that an end marker has arrived from every owner and that every payload it counts has arrived. On object storage, the end markers are manifest entries. In~an attempt that stages, the same arrival check also runs before the request gate closes, and \texttt{flush} runs it again afterward. Queueing acknowledgments from relay-tree hops and the immediate responses to control calls confirm only queueing or registration, so \texttt{flush} also drains the apply queues, synchronizes devices, and collects completion acknowledgments that report the configured post-apply checks. Figure~\ref{fig:acks}, top, follows one path of the relay tree from an owner over the cross-cluster hop, drawn orange, to a rollout node's apply queue, native loader, and receiver storage. Each hop only acknowledges that it queued a payload, and after \texttt{flush}, each receiver sends its completion acknowledgment, the ACK of Figure~\ref{fig:architecture}. A~failed \texttt{flush} returns $(\bot,\mathsf{failed})$, which the coordinator treats as a $\mathsf{failed}$ exception. Each attempt also has a delivery deadline, which its coordinator sets for the attempt's mode and lengthens for each retry. A~wait of Algorithm~\ref{alg:schedule} that runs past it fails the attempt, as when a lost payload stalls application and the full queues hold back the sends before \texttt{flush}. The \texttt{stop\_and\_wait} that follows stops delivery and \mbox{releases the held-back streams}.\nowidowpar

\begin{figure}[!htb]%
\centering
\includegraphics[width=\columnwidth]{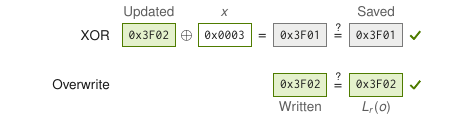}
\caption{Post-apply checks on the example.}
\label{fig:postcheck}
\end{figure}

\paragraph{Post-apply checks.} Optional post-apply checks verify the bits that each intercepted copy writes in place (Figure~\ref{fig:postcheck}). For an XOR copy, the check saves the destination's pre-copy bits and verifies that XORing the mask into the updated bits restores the pre-copy bits. The check reuses the mask that the copy applied, so it verifies the write itself. Whether that mask is correct depends on unchanged scratch bits, a loader property validated in advance. For overwrite copies, the check runs after the item's last intercepted write and compares the bits at active positions with the overwrite values~$o$ transformed by the native \mbox{loader,~$L_r(o)$}.\nowidowpar

\newcommand{\figacks}{%
\begin{figure}[!htbp]%
\centering
\includegraphics[width=\columnwidth]{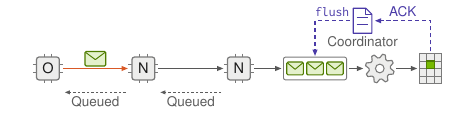}\\[5pt]
\includegraphics[width=\columnwidth]{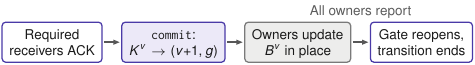}
\caption{Acknowledgments along one relay-tree path (top). The end of a transition (bottom).}
\label{fig:acks}\label{fig:finalize}
\end{figure}}

\subsection{Joint commit and recovery}
\label{app:failure}
\label{app:baseline}

Version~$v$ stays committed through every failure and retry until a joint commit publishes version~$v{+}1$, and rollout never serves partially applied weights. Algorithm~\ref{alg:dcr} runs the attempts of a transition and retries failed ones \mbox{with overwrites}.

\figacks%

\paragraph{Commit and end of the transition.} After \texttt{flush} succeeds with completion acknowledgments from every required receiver, \texttt{commit} replaces $K^v$ with $(v{+}1,g)$ by compare-and-set, which succeeds only if the commit record still equals $K^v$. Owners start the in-place baseline update of \S~\ref{sec:commit} when they read the commit record $(v{+}1,g)$, even when the commit is confirmed late or after failover. Each owner then reports completion to the coordinator. Once every owner has reported (Figure~\ref{fig:finalize}, bottom), the coordinator cleans up and calls \texttt{reopen\_gate}, which invalidates the receivers' KV caches and then reopens the request gate of \mbox{Figure~\ref{fig:pipeline}, bottom}. Rollout resumes with policy version~$v{+}1$, and the coordinator ends \mbox{the transition}.

\begin{algorithm2e}[!t]
\caption{Refit with overwrite recovery.}
\label{alg:dcr}
\DontPrintSemicolon
\Input{$W^{v+1}$, $B^v$, $K^v$, fixed $\mathcal{M}$ and source membership, and the old attempt $\tau_{\mathrm{old}}$ with its candidate-baseline identifier $g_{\mathrm{old}}$ on takeover}
\Output{$v{+}1$ with $\operatorname{bits}(L_r(W^{v+1}))$ on every required receiver, or $\mathsf{reject}$ / $\mathsf{conflict}$}
$m\gets\mathit{mixed}$; $\mathit{admitted}\gets\mathsf{false}$\;
\If{taking over an admitted transition}{
  \texttt{mark\_obsolete}$(\tau_{\mathrm{old}})$; \texttt{stop\_and\_wait}$(\tau_{\mathrm{old}})$\;
  $K\gets\texttt{read\_commit}(\tau_{\mathrm{old}})$\;
  \uIf{$K=(v{+}1,g_{\mathrm{old}})$}{
    \Await baseline update named by $K$\;
    \texttt{cleanup}$(\tau_{\mathrm{old}})$; \texttt{reopen\_gate}$(\tau_{\mathrm{old}})$\;
    \Return{$v{+}1$}\;
  }
  \lElseIf{$K\ne K^v$}{\Return{$\mathsf{conflict}$}}
  \texttt{cleanup}$(\tau_{\mathrm{old}})$\;
  $\mathit{admitted}\gets\mathsf{true}$; $m\gets\mathit{overwrite}$\;
}
\Loop{
  $\tau\gets\bot$; $c\gets\mathsf{failed}$\;
  \Try{
    $g\gets$ new identifier\;
    $\tau\gets\texttt{prepare}(v{+}1,\mathcal{M},g,m)$\;
    $\mathit{admitted}\gets\mathsf{true}$\;
    $(\mathit{ack},\sigma)\gets$ concurrent schedule in Algorithm~\ref{alg:schedule}\;
    \lIf{$\sigma\ne\mathsf{success}$}{\Raise $\mathsf{failed}$}
    $c\gets\texttt{commit}(\tau)$\;
  }
  \Catch{$\mathsf{reject}$}{
    \lIf{$\neg\mathit{admitted}$}{\Return{$\mathsf{reject}$}}
  }
  \lCatch{$\mathsf{conflict}$}{\Return{$\mathsf{conflict}$}}
  \lCatch{$\mathsf{failed}$}{\Pass}
  \If{$c=\mathsf{unknown}$}{
    $K\gets\texttt{read\_commit}(\tau)$\;
    $c\gets\begin{cases}
      \mathsf{success} & K=(v{+}1,g),\\
      \mathsf{failed} & K=K^v,\\
      \mathsf{conflict} & \text{otherwise}.
    \end{cases}$\;
  }
  \If{$c=\mathsf{success}$}{
    \Await baseline update named by $(v{+}1,g)$\;
    \texttt{cleanup}$(\tau)$; \texttt{reopen\_gate}$(\tau)$; \Return{$v{+}1$}\;
  }
  \If{$\tau\ne\bot$}{
    \texttt{mark\_obsolete}$(\tau)$; \texttt{stop\_and\_wait}$(\tau)$\;
    \texttt{cleanup}$(\tau)$\;
    \lIf{$c=\mathsf{conflict}$}{\Return{$\mathsf{conflict}$}}
  }
  \lIf{$\mathit{admitted}$}{$m\gets\mathit{overwrite}$}
}
\end{algorithm2e}

\begin{figure}[!b]%
\centering
\includegraphics[width=\columnwidth]{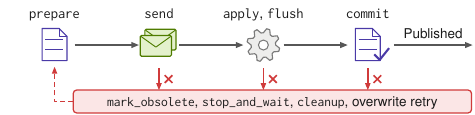}\\[5pt]
\includegraphics[width=\columnwidth]{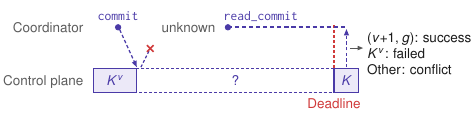}
\caption{The commit sequence and its retry path (top). Resolving an unknown commit outcome after a lost reply (bottom).}
\label{fig:commitseq}\label{fig:readcommit}
\end{figure}

\paragraph{Commit outcomes.} Commit outcome $c$ is $\mathsf{success}$, $\mathsf{failed}$, $\mathsf{unknown}$, or $\mathsf{conflict}$. If the coordinator cannot confirm the outcome, \texttt{commit} returns $\mathsf{unknown}$. Before the coordinator decides whether to retry after an $\mathsf{unknown}$ outcome or a failover, \texttt{read\_commit}$(\tau)$ waits until no commit of~$\tau$ can still change the commit record. Each commit carries a deadline, set by its coordinator, that bounds this wait: the control plane accepts a commit only before its deadline and never after its attempt is marked obsolete (Figure~\ref{fig:readcommit}, bottom). The~commit record $(v{+}1,g)$ confirms success, unchanged $K^v$ confirms failure, and~any other value signals a \mbox{commit} conflict: another coordinator committed the transition first. In Figure~\ref{fig:readcommit}, bottom, the reply to \texttt{commit} is lost, so the coordinator cannot tell whether the commit record holds $K^v$, $(v{+}1,g)$, or another coordinator's commit until \texttt{read\_commit} returns after the deadline. Of these outcomes, only a confirmed failure takes the retry path of Figure~\ref{fig:commitseq}, top, through \texttt{mark\_obsolete}, \texttt{stop\_and\_wait}, and \texttt{cleanup} to an overwrite retry under a new~$\tau$. Because \texttt{flush} has already drained the attempt, \texttt{stop\_and\_wait} on that \mbox{path finds nothing pending}.

\begin{figure}[!t]%
\centering
\includegraphics[width=\columnwidth]{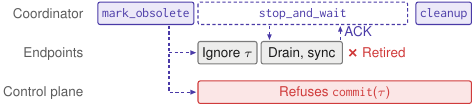}\\[5pt]
\includegraphics[width=\columnwidth]{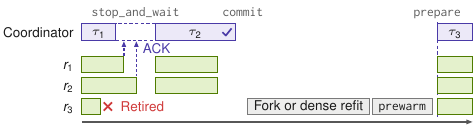}
\caption{The calls that stop a failed attempt~$\tau$ (top). Receiver membership across attempts (bottom).}
\label{fig:stopattempt}\label{fig:membership}
\end{figure}

\paragraph{Stopping an attempt.} Recovery first stops the failed attempt with three calls. After the coordinator calls \texttt{mark\_obsolete}, endpoints ignore the attempt's later payloads and \texttt{open\_apply}, and the control plane refuses the attempt's \texttt{commit} (Figure~\ref{fig:stopattempt}, top). Then \texttt{stop\_and\_wait}, drawn dashed, stops delivery and relay-tree forwarding, waits for pending transport work and receiver writes to \mbox{finish}, synchronizes devices, and collects acknowledgments from all surviving endpoints. Owners finish the stopped attempt's \texttt{build\_complete} without sending, so every training rank leaves its collectives together. A terminated endpoint is retired from the failed attempt only after its process and pending writes are confirmed unable to resume. A receiver whose loader call stays blocked in a cross-rank collective, because a peer lacks the payload or has failed, cannot finish its writes, so \texttt{stop\_and\_wait} terminates and retires it. Retiring a receiver also \mbox{terminates} and retires the other \mbox{receivers} of its serving instance, whose loader collectives need it. Changes in receiver membership take effect only between attempts, so a retired receiver is not a required receiver of the retry (Figure~\ref{fig:membership}, bottom). After \texttt{stop\_and\_wait} returns, \texttt{cleanup} releases the attempt's payloads and delta records, leaving $B^v$ unchanged. A failure before a successful \texttt{prepare} leaves \mbox{nothing to stop or clean up}.

\paragraph{Joining receivers.} A new or restarted receiver restores the committed version, either by forking a receiver that still holds it and is not being written or by a dense refit. It then runs \texttt{prewarm} and joins at the next \texttt{prepare} if that version is still committed (Figure~\ref{fig:membership}, bottom). Otherwise, it first restores the newer committed version the same way. Between attempts of one transition, the surviving receivers may hold partial writes and cannot be forked, so a receiver that joins before the retry restores the committed version by a dense refit from the fixed source baseline~$B^v$. It joins at the retry's \texttt{prepare} as a required receiver, and because the retry overwrites the complete changed set, it reaches $P_r^{v+1}$ like the surviving receivers. Until it joins, it is not a required receiver of any attempt, and its instance serves no requests. The receivers of a serving instance join together. Figure~\ref{fig:membership}, bottom, draws writes in green. Receiver~$r_3$, which serves an instance of its own, fails during~$\tau_1$ and is retired. Receivers $r_1$ and~$r_2$ finish their writes and acknowledge, so the retry~$\tau_2$ requires only them. After a fork or dense refit and \texttt{prewarm}, $r_3$~joins the next transition's \mbox{attempt~$\tau_3$ at \texttt{prepare}}.

\newcommand{\figfailover}{%
\begin{figure}[!htbp]%
\centering
\includegraphics[width=\columnwidth]{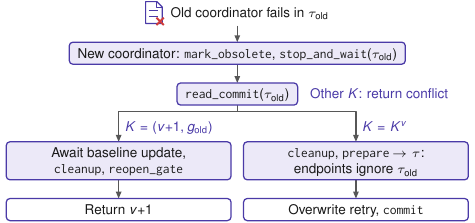}\\[5pt]
\includegraphics[width=\columnwidth]{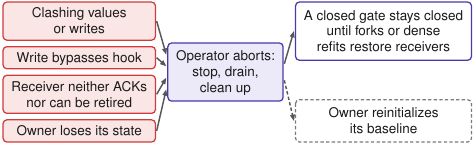}
\caption{Coordinator takeover of an admitted transition (top). Operator intervention for four kinds of persistent failure (bottom).}
\label{fig:failover}\label{fig:operator}
\end{figure}}

\paragraph{Coordinator failover.} A new coordinator uses the control plane to decide whether to start the transition afresh, retry it, or finish it (Figure~\ref{fig:failover}, top). The control plane keeps the attempt of every successful \texttt{prepare} until the transition ends, so a new coordinator learns from it whether the transition was admitted. If not, it has no attempt to stop and starts in mixed mode. If so, it first marks the old coordinator's last attempt $\tau_{\mathrm{old}}$ obsolete, stops it, and drains its writes, then waits with \texttt{read\_commit} until no commit of $\tau_{\mathrm{old}}$ can change the commit record. If such a commit took effect, the new coordinator waits for the baseline update named by the commit record, cleans up, reopens the gate under $\tau_{\mathrm{old}}$, and returns $v{+}1$, because $\tau_{\mathrm{old}}$ committed the same fixed $W^{v+1}$ after every required receiver acknowledged. If the record holds another coordinator's commit, the new coordinator has itself been replaced and returns~$\mathsf{conflict}$. If the commit record remains $K^v$, the new coordinator cleans up and retries in overwrite mode from the \mbox{unchanged} $K^v$, repairing any partial writes before its own commit \mbox{replaces~$K^v$ with~$(v{+}1,g)$}.

\paragraph{Replaced coordinator.} Commit conflicts arise only from coordinator failover, and only a replaced coordinator sees one. Once the new coordinator's \texttt{prepare} replaces $\tau_{\mathrm{old}}$, endpoints ignore whatever the old coordinator still sends. The control plane also raises a $\mathsf{conflict}$ exception for any later \texttt{prepare} or \texttt{commit} of the old coordinator, which then returns without retrying or touching the request gate, \mbox{because} the new coordinator stops and cleans up $\tau_{\mathrm{old}}$. The old coordinator can also learn of a conflict from \texttt{read\_commit} after its own commit, and it then stops its attempt, cleans up, and~returns without retrying, leaving the request gate to the coordinator \mbox{that committed}. If the old \mbox{coordinator's} own commit took effect, it may also clean up and call \texttt{reopen\_gate}, but receivers act on \texttt{reopen\_gate} once per attempt and ignore it for any \mbox{attempt} but their latest, so a late call changes nothing. After \texttt{cleanup} and \texttt{reopen\_gate}, the active coordinator ends the transition by having the control plane drop the transition's attempts, so a takeover before then finds them and reopens the gate. Training takes its next optimizer step only afterward, so a takeover never repeats a baseline update from \mbox{the next step's shards}.

\paragraph{Baseline update after failover.} Owners hold their candidate shards and their baselines in memory, so~they can finish the in-place baseline update after a coordinator failover. \mbox{Because} the update writes absolute values derived from the fixed $W^{v+1}$, an owner that reads the commit record again can safely repeat it. Failover therefore needs no durable log of \mbox{which owners have finished}.

\figfailover%

\newsavebox{\apigrid}
\newlength{\apichar}\newlength{\apileftw}\newlength{\apirightw}\newlength{\apiindent}\newlength{\apigap}
{\ttfamily\codefontsize\global\apichar=0.55em}
\setlength{\apileftw}{41\apichar}
\setlength{\apirightw}{37\apichar}
\setlength{\apiindent}{\dimexpr(\columnwidth-40\apichar)/2+2\apichar-4pt-\fboxrule\relax}
\setlength{\apigap}{\dimexpr\columnwidth+\columnsep-41\apichar\relax}
\begin{lrbox}{\apigrid}
\begin{minipage}{\dimexpr\textwidth-8pt-2\fboxrule\relax}
\hspace*{\apiindent}\begin{minipage}[t]{\apileftw}
\begin{lstlisting}[style=dcrcode]
# Transport adapter: three calls
class ObjectStorage:
    def send(self, tau, p):
        put(key(tau, p.id), p)
        post_manifest(tau, p.id)
    def stop_and_wait(self, tau):
        stop(tau)
        return wait_until_done(tau)
    def cleanup(self, tau):
        delete(tau)
\end{lstlisting}
\end{minipage}\hspace*{\apigap}\begin{minipage}[t]{\apirightw}
\begin{lstlisting}[style=dcrcode]
class RelayTree:
    def send(self, tau, p):
        root.push(tau, p)
    def stop_and_wait(self, tau):
        stop_forwarding(tau)
        return collect_acks(tau)
    def cleanup(self, tau):
        release(tau)
\end{lstlisting}
\end{minipage}\hspace*{\fill}\\[6pt]
\hspace*{\apiindent}\begin{minipage}[t]{\apileftw}
\begin{lstlisting}[style=dcrcode]
# Training side: conversion tasks
for T in conversion_tasks(model):
    track(T)  # task tracker, every task
    if affine(T, M):
        # direct projection
        project(T)
    else:
        # residual conversion
        residual(T)
# Mapping metadata gives shard
# axes and offsets.
# No model-specific placement logic.
\end{lstlisting}
\end{minipage}\hspace*{\apigap}\begin{minipage}[t]{\apirightw}
\begin{lstlisting}[style=dcrcode]
# Rollout side: native loader
prewarm(M)       # scratch, skips
on payload p:    # either transport
    ingest(tau, p.id, p)
def apply(p):
    for name, idx, vals in p.items:
        scratch[:] = q
        scratch[idx] = vals
        with dispatch_hook():
            loader(name, scratch)
# No model-specific placement logic.
\end{lstlisting}
\end{minipage}\hspace*{\fill}
\end{minipage}
\end{lrbox}
\newsavebox{\apilisting}
\begin{lrbox}{\apilisting}
\setlength{\fboxsep}{4pt}%
\fcolorbox[gray]{0.2}{0.85}{\usebox{\apigrid}}%
\end{lrbox}
\newcommand{\apifigure}{\begin{figure*}[!t]
\centering
\usebox{\apilisting}
\caption{Integration sketch: a shared core with three interfaces.}
\label{fig:api}
\end{figure*}}

\paragraph{Operator intervention.} Some persistent failures fall outside the automatic recovery of Algorithm~\ref{alg:dcr} and require operator intervention: clashing canonical values or destination writes, an active value that matches every validated placeholder of its path, a write that bypasses the dispatch hook, a receiver that neither acknowledges nor can be retired, a \texttt{prepare} that keeps raising $\mathsf{reject}$, or an owner that loses its candidate shards or its baseline in a crash.

The operator aborts the transition: the coordinator stops and drains the attempt and cleans up, and a closed request gate stays closed until forks or dense refits restore every required receiver to the committed version (Figure~\ref{fig:operator}, bottom). An~owner that lost its state also reinitializes its baseline at the committed version, as after a source membership change, drawn dashed in the figure. The active coordinator ends the transition once rollout resumes \mbox{with the committed version}.

\section{Integration and Measurement}
\label{app:integration}
\apifigure

Appendix~\ref{app:implementation} covers the implementation behind \S~\ref{sec:design}, the memory accounting for \S\S~\ref{sec:detect} and~\ref{sec:apply-scratch}, and the pipeline settings for \S~\ref{sec:pipeline}. Appendix~\ref{app:measurement} gives measurement details for \S~\ref{sec:evaluation}. The cited NeMo RL commit~\cite{nemorl2026dcrcommit} contains the Megatron Bridge and vLLM integration used in the experiments.

\subsection{Implementation}
\label{app:implementation}

Figure~\ref{fig:api} sketches the integration. The codec, receiver checks, commit, and recovery of Algorithms~\ref{alg:schedule} and~\ref{alg:dcr} form a shared core with three interfaces: a transport adapter, conversion tasks with mapping metadata on the training side, and the serving runtime's native loader with a dispatch hook on its final GPU storage copies. In the sketch, \texttt{M} is the integration setup~$\mathcal{M}$. Each transport adapter implements only \texttt{send}, \texttt{stop\_and\_wait}, and \texttt{cleanup} of Table~\ref{tab:api}. Appendix~\ref{app:sourceconditions} lists the training-side conversion \mbox{tasks and mapping classes}.

\paragraph{Training side.} Owners compare, project, encode, and compress on the CPU, where the source baseline is kept. For each chunk, an owner copies its candidate shards from GPU memory into pinned staging buffers and compares and projects them there. Only residual tasks need change flags, whose all-reduce gives every owner the same $\mathcal{U}_{\mathrm{req}}$, so all owners convert their tensors in the same order. Megatron Bridge converts the residual parts on the GPU and assembles them with its own collectives, which all owners thus enter together. The assembling owner copies the result into pinned host memory, where it compares the result \mbox{with its residual baseline}.

\begin{figure}[!b]%
\centering
\includegraphics[width=\columnwidth]{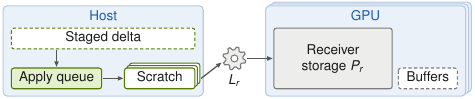}\\[5pt]
\includegraphics[width=\columnwidth]{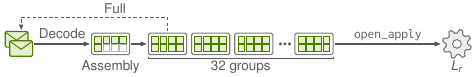}
\caption{Memory of one rollout node beyond receiver storage~$P_r$ (top). The apply queue (bottom).}
\label{fig:rxmem}\label{fig:staging}
\end{figure}

\paragraph{Rollout side.} The rollout side adds three pieces around the native loader of vLLM~0.17.0: a \texttt{prewarm} call, a dispatch hook, and an apply queue. For \texttt{prewarm}, receivers obtain canonical names, shapes, and dtypes from the training side. The~dispatch hook intercepts the loader's final GPU storage copies, which are \texttt{aten.copy\_} operations. The apply queue holds up to 32 groups of at most eight decoded payloads, and a separate assembly group holds at most eight more, so the queue and its assembly group hold at most 264 decoded payloads per rollout node (Figure~\ref{fig:staging}, bottom). In synchronous RL, arriving payloads are decoded into the assembly group, which then joins the queue. After \texttt{open\_apply}, receivers apply payloads from the queue in declared order and take the next one from the assembly group if it is not yet queued. A full queue holds back ingestion and delivery. In asynchronous RL, the staged delta feeds the queue after \texttt{open\_apply}. Figure~\ref{fig:rxmem}, top, places the queue and the staged delta in a rollout node's memory beyond receiver storage~$P_r$. The node's receivers share both. Each receiver has its own scratch and GPU, drawn stacked, and its native loader~$L_r$ copies scratch into~$P_r$. Dashed boxes are temporary: the staged delta and GPU buffers \mbox{for masks and post-apply checks}.

Scratch, the queue, and the staged delta live in host memory, with scratch pinned for direct copies to GPU memory. The loader's final copies move each receiver's slice of scratch to the GPU, where the dispatch hook XORs masks into receiver storage or writes active values under the active mask, using temporary GPU buffers for XOR masks, active masks, and~\mbox{optional post-apply checks}.\nowidowpar

\paragraph{Memory accounting.} Table~\ref{tab:memory-accounting} reports the host memory that \sysname{} adds per node for both Qwen3 models with real updates of \S~\ref{sec:eval-crossstudy} on the \S~\ref{sec:eval-setup} latency testbed. A training node runs four ranks, and a rollout node runs eight receivers. The table excludes receiver storage, pinned staging buffers, held-back payloads, and unrelated process memory. The source baseline of \S~\ref{sec:detect} holds one copy of the source weights plus a residual baseline of 2.25~GB for the 61.06~GB \mbox{Qwen3-30B-A3B} checkpoint and 9.59~GB for the 470.2~GB \mbox{Qwen3-235B-A22B} one. Across the eight training nodes, the source baseline takes 7.91~GB per node at 30B and 59.97~GB at 235B\@. Held-back payloads stay compressed in host memory until they are sent or the \texttt{cleanup} of their attempt releases them, so an owner holds back at most its \mbox{payloads of one attempt}.

On a rollout node, the eight receivers share one apply queue and its assembly group, which hold the whole 30B delta but at most 264 of the decoded 235B payloads. The receivers take payloads from the queue independently, so each has its own scratch, sized for the largest canonical tensor: the embedding or output head, 0.622~GB at 30B and 1.245~GB at 235B (Figure~\ref{fig:rxmem}, top). Each receiver fills its scratch with whole items, because only its native loader selects the receiver's slice. A checkpoint that stacks a layer's experts in one tensor thus needs scratch as large as that \mbox{stacked tensor on every receiver}.

In asynchronous RL, each rollout node instead stages one decoded copy of the delta, which its eight receivers share until application, and the queue draws its payloads from that copy. Staging decoded payloads keeps decompression out of the pause while the gate is closed. A retry that starts with the gate closed stages nothing, and one that starts with the gate open stages as many decoded bytes as the first attempt, because it recomputes the same changed set and an overwrite value is as wide as \mbox{an XOR mask}. Unlike the bounded apply queue, the staged delta grows with the changed set, so staging needs host memory for one decoded delta per rollout node. With the real updates of \S~\ref{sec:eval-crossstudy}, this delta takes about 4\% of the checkpoint at both Qwen3 sizes (Table~\ref{tab:memory-accounting}), which would be about 89~GB per rollout node for the 2{,}215~GB 1T checkpoint. Deployments~that lack this memory can still refit as in synchronous RL, streaming through the bounded apply queue while rollout pauses for \mbox{the whole refit}. The synchronous rollout total sums the queue and scratch rows, and the asynchronous one sums the scratch and staged-delta rows. \sysname{} adds no model-sized GPU allocation, because receivers update their storage in place and keep only temporary GPU buffers for \mbox{masks and optional post-apply checks}.

\begin{table}[!htbp]%
\caption{Host memory that \sysname{} adds per node with real updates.}
\label{tab:memory-accounting}
\centering
\footnotesize
\setlength{\tabcolsep}{3pt}
\begin{tabular}{@{}>{\raggedright\arraybackslash}p{0.57\columnwidth}rr@{}}
\toprule
Component & \shortstack{Qwen3\\30B-A3B\\(GB)} & \shortstack{Qwen3\\235B-A22B\\(GB)} \\
\midrule
Source baseline, per training node & 7.91 & 59.97 \\
\midrule
Apply queue, synchronous RL & 2.47 & 5.67 \\
Scratch & 4.98 & 9.96 \\
Staged delta,\newline asynchronous RL only & 2.47 & 18.81 \\
\midrule
Rollout total, synchronous RL & 7.45 & 15.63 \\
Rollout total, asynchronous RL & 7.45 & 28.77 \\
\bottomrule
\end{tabular}
\end{table}

\paragraph{Pipeline settings.} The chunk targets of \S~\ref{sec:eval-setup} gave each transport the lowest refit latency in every configuration. These targets and the 512~MiB bucket target count input bytes before unchanged elements are dropped: bytes of local source tensors for affine tasks and of converted canonical tensors for residual tasks. A~bucket counts the full input size of each chunk that emits records, so buckets hold similar comparison work, while their payload sizes vary with the change rate. Tensors and chunks are never split, so each owner emits at most one item per canonical tensor, and a chunk or bucket can exceed its target. Owners copy and compare one chunk at a time, so larger chunks pay the per-chunk overhead less often, while smaller ones shrink the pinned staging buffers. Because chunks are never split, smaller chunks also overshoot the bucket target less. Figure~\ref{fig:chunkbucket} redraws the chunks of Figure~\ref{fig:pipeline}, top, by input size, with dashed lines at the 256~MiB relay-tree chunk target and the bucket target. Bucket~$\ell$ adds up the full input sizes of its two chunks with records and skips the unchanged chunk~($\varnothing$). The tall tensor exceeds both targets, so its chunk and bucket~$\ell{+}1$ do too. The~different chunk targets of the two transports shift bucket boundaries, so their payload volumes differ slightly in Table~\ref{tab:latency}, although both yield 482 payloads \mbox{across 32 owners at 120B}.\nowidowpar

\newcommand{\figchunkbucket}{%
\begin{figure}[!htbp]%
\centering
\includegraphics[width=\columnwidth]{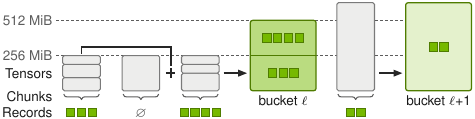}
\caption{Chunk and bucket sizes in input bytes.}
\label{fig:chunkbucket}
\end{figure}}
\figchunkbucket%

\newcommand{\figcodecbench}{%
\begin{figure}[!t]%
\centering
\includegraphics[width=\columnwidth]{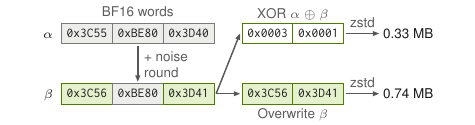}
\caption{Synthetic codec benchmark at $a=10^{-4}$.}
\label{fig:codecbench}
\end{figure}}

\subsection{Measurement}
\label{app:measurement}

\begin{table}[!htbp]%
\caption{Checkpoints used in \S\S~\ref{sec:eval-scale} and~\ref{sec:eval-floors}.}
\label{tab:latmodels}
\centering
\footnotesize
\renewcommand{\arraystretch}{1.15}
\setlength{\tabcolsep}{4pt}
\begin{tabular}{@{}l>{\raggedright\arraybackslash}p{0.62\columnwidth}r@{}}
\toprule
Size & Checkpoint & GB \\
\midrule
30B & Nemotron-3-Nano-30B-A3B & 63.2 \\
120B & Nemotron-3-Super-120B-A12B & 247.2 \\
235B & Qwen3-235B-A22B & 470.2 \\
550B & Nemotron-3-Ultra-550B-A55B & 1{,}121 \\
1T & Nemotron-3-Ultra-550B-A55B with twice the layers & 2{,}215 \\
\bottomrule
\end{tabular}
\end{table}

\begin{figure}[!t]%
\centering
\includegraphics[width=\columnwidth]{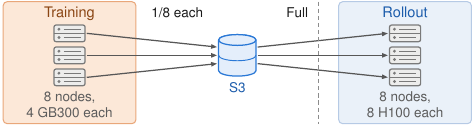}\\[5pt]
\includegraphics[width=\columnwidth]{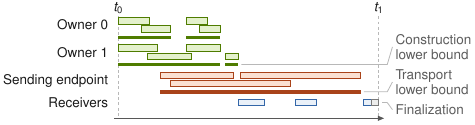}
\caption{The transport-only full-checkpoint reference on the latency testbed (top). The timed window and the traced intervals of one relay-tree refit (bottom).}
\label{fig:testbed}\label{fig:window}
\end{figure}

\paragraph{Latency runs.} Table~\ref{tab:latmodels} lists the checkpoints of the latency runs. Figure~\ref{fig:testbed}, top, shows their testbed, drawing three of the eight nodes on each side, and the transport-only full-checkpoint reference of \S~\ref{sec:eval-setup}, which moves the checkpoint through S3 to every rollout node across the dashed cluster boundary. \mbox{Nemotron-3-Nano-30B-A3B} is the 30B checkpoint of these runs, whereas the system-comparison, ablation, and training runs of \S\S~\ref{sec:eval-crossstudy}, \ref{sec:eval-ablation}, and~\ref{sec:eval-fidelity} use \mbox{Qwen3-30B-A3B} as their 30B model. In the 3\% and 5\% stress cases, uniformly chosen changes reach every residual task, so $\mathcal{U}_{\mathrm{req}}$ \mbox{holds every residual tensor}.\nowidowpar

\paragraph{Timed window.} Each refit's timed window runs from $t_0$ to $t_1$ (Figure~\ref{fig:window}, bottom). At~$t_0$, the first owner begins delta construction, and at $t_1$, the last owner finishes its baseline update after the joint commit. Finalization runs from the end of the last loader application to~$t_1$. Refit \mbox{latency} is $t_1-t_0$ minus the time of the post-apply checks, which verify the written bits before the ACK\@. It also excludes benchmark change injection and the receivers' \texttt{prewarm}. The \texttt{prepare} call precedes~$t_0$, and \texttt{cleanup} and the gate reopening follow~$t_1$. The latencies of ByteCheckpoint, \mbox{AReaL-DTE}, and PULSE in Table~\ref{tab:crossstudy} include each system's own checkpoint save and load, which the \mbox{transport-only reference excludes}.\nowidowpar

\paragraph{Lower bounds.} Each lower bound of \S~\ref{sec:eval-floors} is the longest union of traced intervals. In Figure~\ref{fig:window}, bottom, outlined boxes are intervals, and solid strips are their unions, which count overlaps once and exclude gaps. The longest of the owners' unions gives the construction lower bound, and the longest of their sending endpoints' unions, one of which is drawn, gives the transport lower bound. The receivers row shows the loader applications, and the gray box is finalization, which neither bound covers. Each~\mbox{construction} interval runs from an operation's start until its output is ready: change flags or locations for comparison, mapped indices and values for \mbox{projection}, canonical \mbox{tensors} for residual conversion, location and value buffers for encoding, and compressed bytes for zstd. For non-blocking operations, the interval ends at completion, not when the call returns. Each relay-tree transport interval runs from payload submission until the owner's sending endpoint \mbox{receives} the root's acknowledgment that it queued the payload. In~the relay-tree latency runs, the apply queue never filled, so~\mbox{no interval waited for queue space}. Before taking the unions, we map Nsight Systems trace timestamps onto a common time axis and clip each interval to $[t_0,t_1]$, \mbox{discarding} empty intersections. We average per-refit bounds as \S~\ref{sec:eval-setup} averages refit latency. Each~union ends before finalization, and mean refit latency exceeds both mean bounds at every traced checkpoint size (Figure~\ref{fig:floors}). At 235B, the \mbox{construction} bounds of 15.4 and 18.6~s stay below the standalone construction times of 20.23 and \mbox{23.07~s in Table~\ref{tab:bridgeablation}}.

\begin{table}[!htbp]%
\caption{Compressed sizes (MB) of the changed values of 20 million synthetic BF16 inputs.}
\label{tab:compression}
\centering
\footnotesize
\setlength{\tabcolsep}{2.5pt}
\begin{tabular}{@{}cccc@{}}
\toprule
Noise $a$ & Overwrite $\downarrow$ & XOR $\downarrow$ & Reduction $\uparrow$ \\
\midrule
$10^{-4}$ & \phantom{0}0.74 & \phantom{0}0.33 & 2.21$\times$ \\
$10^{-3}$ & \phantom{0}7.02 & \phantom{0}3.23 & 2.18$\times$ \\
$10^{-2}$ & 25.70 & 14.93 & 1.72$\times$ \\
\bottomrule
\end{tabular}
\end{table}

\paragraph{Synthetic codec benchmark.}\phantomsection\label{app:codec-benchmark} The benchmark of \S~\ref{sec:eval-ablation}, with the sizes of Table~\ref{tab:compression}, runs PyTorch~2.14.0 and NumPy~2.4.6 on four CPU threads, with random seed 42 for the BF16 inputs and 44 for the FP32 noise. The same input and noise vectors serve all three values of $a$, which change 469{,}195, 4{,}487{,}140, and 16{,}516{,}798 of the 20 million inputs. Scaling~the noise by $a$ and adding it to the input are separate FP32 operations. Rounding to BF16 turns the inputs~$\alpha$ into~$\beta$ before the bitwise comparison that selects the changed words for the XOR and overwrite streams (Figure~\ref{fig:codecbench}). Table~\ref{tab:compression} gives the stream sizes in MB of $10^6$ bytes and reductions computed from exact byte counts. Each~stream holds only little-endian 16-bit words compressed with zstd~1.5.5 at level~1. The~overwrite stream decodes to the updated BF16 words, and XORing the decoded masks into the matching input words \mbox{gives the same result}.\nowidowpar

\figcodecbench

\section{GRPO Experiment Settings}
\label{app:probes}

This appendix details the GRPO settings for the element-change measurements of \S~\ref{sec:sparsity}, the system comparison of \S~\ref{sec:eval-crossstudy}, and the training runs of \S~\ref{sec:eval-fidelity}.

\newcommand{\figkills}{%
\begin{figure}[!ht]%
\centering
\includegraphics[width=\columnwidth]{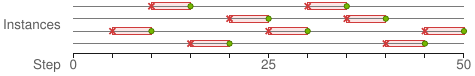}
\caption{Receiver-kill schedule of the \sysname{} runs of \S~\ref{sec:eval-fidelity}.}
\label{fig:kills}
\end{figure}}

\paragraph{Element-change measurements.} These measurements follow the first stage of the agentic SWE RL recipe for Qwen3~\cite{nemorl2026sweqwen}, with the data and GRPO settings of the runs below. The six measured models are \mbox{Qwen3-30B-A3B}, Llama-3.2-3B, \mbox{Gemma-3-4B}, and the 0.5B, 1.5B, and~7B Qwen2.5 models. The element change rate varies across runs and steps: for \mbox{Qwen3-30B-A3B}, it is 1.2\% per step over the first five steps of these measurements, about 2\% in steps 6--15 of the \S~\ref{sec:eval-crossstudy} runs, and 1.2\% on average over the 50 asynchronous steps \mbox{of \S~\ref{sec:eval-fidelity}}.

\figkills

\paragraph{System-comparison and training runs.} In both \S\S~\ref{sec:eval-crossstudy} and~\ref{sec:eval-fidelity}, the runs use the first stage of the same recipe, with the Nemotron-RL-Super-Training-Blends dataset file \texttt{swe1.jsonl}, BF16 weights, FP32 AdamW state, and the recipe's constant learning rate of $10^{-6}$ with no warmup, weight decay, or KL penalty. This stage rewards single-step tool calls that match the arguments of the labeled action. Each GRPO step uses 64 prompts with 8~responses each and a batch size of 512. Clipping bounds of 0.2 below and 0.28 above limit probability ratios to $[0.8,\,1.28]$. Figure~\ref{fig:training}, bottom, plots the per-token KL between rollout and training policies, logged as \texttt{train/gen\_kl\_error}. The training runs of \S~\ref{sec:eval-fidelity} are asynchronous, with trajectories at most one training step old. The \sysname{} runs kill receivers on the schedule of Figure~\ref{fig:kills}, which marks each kill with a cross and each restart five steps later with a dot for an illustrative choice of instances. No other failure of Table~\ref{tab:failures} \mbox{is injected}.

\section{Related Systems}
\label{app:comparison}

For each requirement G1--G6 of \S~\ref{sec:goals}, this appendix explains how the systems of Table~\ref{tab:goals} meet it according to their cited sources, and then describes the other related systems.

\paragraph{\sysname{}.} The design meets all six requirements. For~G1, placement stays in the native loader with no model-specific placement logic, as \S\S~\ref{sec:apply-scratch} and~\ref{sec:eval-setup} describe. Section~\ref{sec:guarantee} and Appendix~\ref{app:proof} establish G2, and the bitwise comparisons with dense refits in \S~\ref{sec:eval-fidelity} confirm it. For G3, the payload volumes of Tables~\ref{tab:latency} and~\ref{tab:bridgeablation} scale with the changed set. For G4, both transports deliver without a cross-cluster collective, as \S~\ref{sec:protocol} describes, and the latency runs of \S~\ref{sec:eval-setup} use them between clusters that share no InfiniBand or EFA route. For G5, \S~\ref{sec:eval-floors} shows delivery overlapping delta construction. Payload deduplication, overwrite recovery, and joint commit meet G6, as \S\S~\ref{sec:codec} and~\ref{sec:commit} and Appendix~\ref{app:api} describe, and the receiver kills of \S~\ref{sec:eval-fidelity} \mbox{exercise the last two}.

\paragraph{Native (G1).}
In checkpoint-engine, verl, SparseRL-Sync, and slime, the serving runtime places all weights, and in NCCL-Reshard and TensorHub it performs only part of the placement. The slime framework reloads its patched checkpoint through the native loader. The serving runtime of checkpoint-engine also chooses which weights to copy. \mbox{NCCL-Reshard} copies feed-forward shards directly into vLLM parameters using its own metadata and hooks and passes only the remaining parameters through the native loader. TensorHub copies its published buffers, without interpreting their layout, into tensors that the serving runtime registers, so the runtime defines the destinations. The weights reach these buffers already converted to the rollout layout, outside the native \mbox{loader~\cite{ye2026tensorhub}}. ByteCheckpoint places weights with per-framework load planners, and its description covers \mbox{no serving runtime~\cite{wan2025bytecheckpoint}}.

The other delta systems place updates themselves: ROSE routes them to receiver shards, AuroraRL writes directly into parameter storage, and PULSE writes transmitted values into receiver tensors in the sender's layout. Across clusters, AReaL-DTE remaps updates into rollout layouts at the receiver and falls back to the native loader, which thus performs only part \mbox{of its placement}.

\paragraph{Bit-exact (G2).}
Approaches A, B, and E fully meet G2, and approach D avoids arithmetic reconstruction but meets G2 only partly. NCCL-Reshard and checkpoint-engine send every weight, so they preserve the target bits~\cite{nemorl2026reshard,moonshot2026checkpointengine}. Likewise, \mbox{ByteCheckpoint} saves every tensor shard and, on load, copies the saved byte ranges that overlap each target shard~\cite{wan2025bytecheckpoint}. \mbox{TensorHub copies} its published buffers into the serving runtime's registered tensors, which yields exact target bits because the published buffers hold the weights in the layout and dtype of the registered tensors~\cite{ye2026tensorhub}. The slime framework patches a full local checkpoint with byte-level XOR or overwrite deltas, verifies a checksum of every patched tensor, and reloads the checkpoint through the native loader, which thus writes the same bits as \mbox{a dense refit}~\cite{thudm2026slime}.

No approach-D system states the loader conditions under which its exactness holds. PULSE's Appendix~H.6 proves exact reconstruction of a tensor from a correct base~\cite{miahi2026pulse}, but extending the proof through a native loader requires conditions on the loader's transformations and storage writes. In~verl, NaN masks in intercepted copies protect unchanged positions, and~a~round-trip test verifies that the model's generated outputs are bit-identical~\cite{verl2026deltasharded}, but the loader's supported transformations are not stated. \mbox{SparseRL-Sync} replaces the loader's copy with a NaN-masked sparse scatter and reports tensor-by-tensor bitwise agreement with a dense refit, but does not specify supported transformations or interception of every storage write~\cite{hu2026sparserlsync}. Partial G2 marks thus flag guarantees that these sources leave unstated.

In approach C, arithmetic reconstruction of updated or pre-update values can introduce rounding errors. ROSE sends arithmetic differences~\cite{gao2026rose}. AuroraRL claims bit-exact updates from scatter-added deltas but states no conditions under which these additions reproduce the target bits~\cite{ruan2026aurorarl}. AReaL-DTE~\cite{peng2026arealdte} reconstructs old weights by inverting AdamW and compares BF16 bit \mbox{patterns} after aligning them to a common layout. Its~authors note rounding in fused AdamW implementations, so~comparison against these reconstructed old weights can miss a changed element.

\paragraph{Delta (G3).}
Approaches A and B move every weight byte, whereas approaches C, D, and E scale with the set of changed values. In approach D, PULSE, \mbox{SparseRL-Sync}, and verl send absolute overwrites as indices and BF16 values of changed elements. The verl framework detects changes on each training rank's shard and maps them to full-tensor coordinates by block placement, but supports only conversions that permute elements~\cite{verl2026deltaep,verl2026deltasharded}. AReaL-DTE sends target values of detected changes, AuroraRL packages each step as a BF16 delta checkpoint, and slime compresses its deltas with zstd, which compresses away the zero bytes that XOR \mbox{leaves at unchanged elements}.\nowidowpar

\paragraph{Decoupled (G4).}
NCCL-Reshard, SparseRL-Sync, and verl each use a collective that spans training and rollout. \mbox{The checkpoint-engine} system runs alongside serving runtimes, loads full weights onto GPUs, and broadcasts them to its workers. Its description mentions no cross-cluster transport~\cite{moonshot2026checkpointengine}. TensorHub, ByteCheckpoint, ROSE, AReaL-DTE, AuroraRL, PULSE, and slime move weights through replication, relays, or shared storage without a \mbox{cross-cluster} collective. TensorHub places one full-weight replica in each remote datacenter before replicating locally. AuroraRL~streams each delta checkpoint through \mbox{one seed actor per region}.

\paragraph{Pipelined (G5).}
Delivery overlaps construction or application in ByteCheckpoint, ROSE, AuroraRL, and verl. \mbox{ByteCheckpoint} pipelines serialization with uploads on save and file reads with deserialization, host-to-device copies, and inter-GPU transfers on load~\cite{wan2025bytecheckpoint}. ROSE pipelines bucket publishing with pulling through its relay~\cite{gao2026rose}. AuroraRL emits each segment as soon as it is encoded, so extraction, cross-cluster transfer, and relay fanout overlap in time~\cite{ruan2026aurorarl}. Its actors also stage each delta while they keep generating, then apply it in place at the next batch boundary, as \sysname{} does in asynchronous RL\@. The verl framework~\cite{verl2026deltasharded} streams changes in bucket-sized batches, which receivers apply \mbox{as they arrive}.

NCCL-Reshard, checkpoint-engine, and PULSE overlap only part of a refit. Within each training pipeline stage, \mbox{NCCL-Reshard} sends and places the feed-forward shards in turn, and only the reshards of different stages overlap. Its double-buffered broadcast of the other parameters overlaps conversion and loading with transfer~\cite{nemorl2026reshard}. The \mbox{checkpoint-engine} system broadcasts each bucket into one of two buffers while the serving runtime reloads the previous bucket from the other, but it does not overlap host-to-device copies with broadcasts~\cite{moonshot2026checkpointengine}. PULSE overlaps downloads with applications only when a receiver replays missed versions, and otherwise encodes, uploads, downloads, decodes, applies, and verifies each delta \mbox{in sequence without overlap}~\cite{miahi2026pulse}.

TensorHub, AReaL-DTE, and SparseRL-Sync construct the whole update before delivering it, and slime runs construction, delivery, and reloading in sequence. TensorHub publishes weights only after the trainer converts all of them to the rollout layout, and its pipelined replication overlaps only replication hops~\cite{ye2026tensorhub}. Within a cluster, \mbox{AReaL-DTE} exchanges the change counts of every operation before any payload, and across clusters it commits a version manifest only after all ranks finish~\cite{peng2026arealdte}. \mbox{SparseRL-Sync} packs the changes of every parameter into one message before broadcasting it~\cite{hu2026sparserlsync}. In slime, rollout nodes read a version only after all of its files are \mbox{written}, and engines reload only after every node holds a verified checkpoint, so delivery overlaps neither construction \mbox{nor reloading~\cite{thudm2026slime}}.\nowidowpar

\paragraph{Recovery (G6).}
No existing system in Table~\ref{tab:goals} fully meets G6\@. ROSE, SparseRL-Sync, and verl describe no refit recovery protocol, although verl does checksum each broadcast payload. For \mbox{NCCL-Reshard} and checkpoint-engine, repeating a dense refit can restore weights, but the cited descriptions do not specify a complete protocol for partial failures and policy commit. TensorHub provides transactions, invalidates incomplete replicas, and recovers transfers from another source~\cite{ye2026tensorhub}, but it coordinates version selection only within each \mbox{model-parallel} group. \mbox{ByteCheckpoint makes} each save or load atomic across the workers of a job with an asynchronous barrier and checks checkpoint integrity~\cite{wan2025bytecheckpoint}, but describes neither repair after a worker fails mid-load nor a commit across receivers. \mbox{AReaL-DTE}~commits a version manifest and leaves partially applied versions to its control plane, without describing their repair. AuroraRL checks the version a delta applies to, uses leases for fault handling, stages deltas before activation, and~lets lagging actors replay the delta chain, but does not describe repair after partial writes. The slime framework applies a delta only to the version it was computed from, publishes a full checkpoint after trainer recovery or engine replacement, and can reapply overwrite deltas to finish a partial patch. Its trainer advances the next delta's baseline when it publishes a version, before any \mbox{rollout node applies it}.\nowidowpar

PULSE receivers advance independently, with no joint commit of the source baseline and the receivers' policy version. PULSE's Algorithm~5 checks the local version before decoding~\cite{miahi2026pulse}, and a receiver applies consecutive deltas directly. PULSE also keeps periodic full checkpoints as anchors. After missed versions, the receiver restores a ready anchor and replays later deltas in order. Ready markers prevent use of incomplete uploads, and a full-state hash checks every reconstruction before returning the new weights. A mismatch triggers anchor recovery. Absolute overwrites can be repeated safely and anchor restoration replaces partially applied state, but without a joint commit, PULSE \mbox{meets G6 only partly}.

\paragraph{Other systems.} The other cited systems distribute full weights or deltas, recover failed roles, or compress weights. The INTELLECT-2 run~\cite{jaghouar2025intellect2} relays checksummed full checkpoints in pipelined shards through \mbox{SHARDCAST}. Laminar and ECHO-2 also relay full weights through the rollout fleet. Laminar's trainer pushes them to one master relay, which forwards them along a pipelined chain of relays that keep them in host memory for rollouts to fetch~\cite{sheng2026laminar}. ECHO-2 streams full snapshots along parallel chains of rollout workers, which keep generating on the installed snapshot~\cite{song2026echo2}. As in slime, trainers in Composer~2 and Fireworks upload per-step deltas to shared storage for rollout clusters to rebuild full weights~\cite{cursor2026composer,fireworks2026composer}. A delta sync \mbox{prototype} for TRL~\cite{huggingface2026trldelta} likewise rebuilds full tensors from a snapshot in CPU memory. In vLLM~\cite{vllm2026sparse}, sparse NCCL weight transfer broadcasts overwrites computed by an RL framework and applies them in place through the native loader. It leaves change detection and the refit lifecycle to that framework and is not transactional, so callers must recover from partial updates. For recovery, RobustRL restarts only a failed training or rollout role~\cite{chen2026robustrl}. When~a trainer failure interrupts a rollout's weight pull, the rollout clears its partially updated weights and pulls them again in full. For checkpoint storage, ZipLLM's BitX~\cite{wang2026zipllm} compresses the XOR of each fine-tuned model's weights with those of \mbox{its base model}. UCCL-Zip~\cite{ma2026ucclzip} compresses full BF16 tensors losslessly inside GPU communication, which speeds up dense refits by up to \mbox{47.5\% for a single tensor}.

In concurrent work, verl extended its change detection to Megatron Bridge parameter mappings, running each mapping's conversion on NaN-marked copies to find final Hugging Face coordinates~\cite{verl2026deltamegatron}. Fireworks and Baseten described cross-cluster refits with XOR or overwrite deltas~\cite{fireworks2026arcv3,stefanopoulos2026baseten}. Fireworks rebuilds the full model in each rollout region and checks it against a checksum of the trainer's weights. Baseten applies overwrites in place to the weights on each GPU but states no loader conditions that guarantee dense-refit bits. Neither company describes a joint commit that binds the policy version to the baseline of the next delta. WeightBridge traces each checkpoint element through the native loader to find its rollout destinations, but sends one full copy of the weights per update and leaves fault tolerance to \mbox{future work~\cite{jiang2026weightbridge}}.

\section{End-to-End Estimates}
\label{app:estimates}

This appendix derives two estimates of the main text: the Kimi~K2 step speedup of \S~\ref{sec:eval-scale} and the bandwidth \mbox{limit of \S~\ref{sec:limitations}}.

\newcommand{\tabcrossover}{%
\begin{table}[!htbp]%
\caption{Relay-tree refits and crossover bandwidth.}
\label{tab:crossover}
\centering
\footnotesize
\setlength{\tabcolsep}{3pt}
\begin{tabular}{@{}llrrrr@{}}
\toprule
Size & Change rate & Latency & Transport & Construction & Gbps \\
\midrule
30B & 3\% & 9.8 & 7.7 & 2.4 & 211 \\
120B & 3\% & 22.6 & 18.6 & 8.4 & 235 \\
235B & 3\% & 44.8 & 41.3 & 15.4 & 244 \\
550B & 3\% & 91.9 & 86.0 & 41.8 & 215 \\
\midrule
30B & 5\% & 12.8 & 9.9 & 2.3 & 220 \\
120B & 5\% & 41.6 & 38.7 & 8.8 & 225 \\
235B & 5\% & 77.4 & 72.4 & 18.6 & 202 \\
550B & 5\% & 146.5 & 136.8 & 51.0 & 176 \\
\bottomrule
\end{tabular}
\end{table}}
\tabcrossover%

The step speedup uses the Kimi K2 measurements of Seer~\cite{qin2026seer}. Before Seer's rollout optimizations, rollout, training, and the weight update take 87\%, 10\%, and 3\% of an RL step, and rollout averages \mbox{12{,}202~s} per step, so training takes about 23~min and the update 7~min. Seer shortens rollout to \mbox{8{,}267~s}, 82\% of its step. A step thus takes about \mbox{2.8--3.9~h}, or \mbox{161--227~min} without the update. With a cross-cluster 1T refit in place of that update, a synchronous step adds 87.5~min with the reference of Figure~\ref{fig:fullckpt}, top, or \mbox{150--300~s} with \sysname{}, a \mbox{1.36--1.52$\times$} speedup. In asynchronous RL, training plus a refit takes \mbox{26--28~min} with \sysname{} and about 111~min with the reference, less than the \mbox{138--203~min} rollout phase, so for steps this long, \mbox{both refits overlap rollout}.\nowidowpar

The bandwidth limit uses the construction lower bound with 32 training GPUs. For the 30B--550B relay-tree refits of Figure~\ref{fig:floors}, this bound is 2.3--51.0~s. A dense refit that crosses the cluster boundary once moves the \mbox{63.2--1{,}121~GB} checkpoints faster than this bound only above an aggregate cross-cluster bandwidth of \mbox{176--244~Gbps}, lowest for 550B at 5\% and highest for 235B at 3\%. Table~\ref{tab:crossover} lists this crossover bandwidth in Gbps with the mean latency and both lower bounds of each refit in seconds. A dense refit through the single relay-tree root takes at least the checkpoint size divided by 5~Gbps, from 101~s at 30B to 59~min at 1T, \mbox{7.9--23.6$\times$} longer than the relay-tree refits of Figure~\ref{fig:fullckpt}, bottom. Striping a dense refit over all eight node pairs at the 5~Gbps per-node peak of \S~\ref{sec:eval-setup} reaches 40~Gbps, below \mbox{the 176--244~Gbps crossover}.

\end{document}